\documentclass[letter,journal]{IEEEtran}
\usepackage[]{subcaption}
\usepackage{cite}
\usepackage{amsmath,amssymb,amsfonts}

\usepackage{algorithmicx}
\usepackage[noend]{algpseudocode}
\usepackage{graphicx}
\usepackage{textcomp}
\usepackage{xcolor}

\usepackage{algorithm}
\usepackage{tabularx}
\usepackage{makecell}
\usepackage{easyReview}

\def\BibTeX{{\rm B\kern-.05em{\sc i\kern-.025em b}\kern-.08em
    T\kern-.1667em\lower.7ex\hbox{E}\kern-.125emX}}

\begin{document}

\title{PHandover: Parallel Handover in Mobile Satellite Network \\
\thanks{Portions of this work were presented at the INFOCOM 2024, “Accelerating Handover in Mobile Satellite Network" \cite{infocom24handover}. 

Jiasheng Wu, Shaojie Su, Wenjun Zhu, Xiong Wang, Jingjing Zhang, Xingqiu He, Yue Gao are with Institute of Space Internet, Fudan University and School of Computer Science, Fudan University. 
The corresponding author is Yue Gao. (E-mail: gao.yue@fudan.edu.cn)
}
}

\author{Jiasheng Wu, Shaojie Su, Wenjun Zhu, Xiong Wang, Jingjing Zhang, Xingqiu He, Yue Gao  \\ Fudan University, China \\}

\maketitle
\thispagestyle{empty}
\setreviewsoff
\begin{abstract} 
The construction of Low Earth Orbit satellite constellations has recently spurred tremendous attention from both academia and industry. 5G and 6G standards have specified the LEO satellite network as a key component of the mobile network. However, due to the satellites' fast traveling speed, ground terminals usually experience frequent and high-latency handover, which significantly deteriorates the performance of latency-sensitive applications. To address this challenge, we propose a parallel handover mechanism for the mobile satellite network which can considerably reduce the handover latency. The main idea is to use plan-based handovers instead of measurement-based handovers to avoid interactions between the access and core networks, hence eliminating the significant time overhead in the traditional handover procedure. Specifically, we introduce a novel network function named Satellite Synchronized Function (SSF), which is designed for being compliant with the standard 5G core network. 
Moreover, we propose a machine learning model for signal strength prediction, coupled with an efficient handover scheduling algorithm. We have conducted extensive experiments and results demonstrate that our proposed handover scheme can considerably reduce the handover latency by 21$\times$ compared to the standard NTN handover scheme and two other existing handover schemes, along with significant improvements in network stability and user-level performance.

\end{abstract}


\section{Introduction}

\IEEEPARstart{T}{he} recent emergence of Non-terrestrial-network (NTN) Internet services provided by Low Earth Orbit (LEO) satellites, such as Starlink, has gained wide attention \cite{starlink,lin2024fedsn,peng2025sigchord,yuan2024satsense,zhao2024leo,lin2025leo}. The LEO satellite-based connectivity offers global Internet coverage, serving as a valuable complement to traditional terrestrial networks~\cite{yuan2023graph,peng2024sums,zhang2024satfed,yuan2025constructing}. In this context, 3GPP 5G and 6G standards emphasize the crucial role of satellite communication within the entire system by building the mobile satellite network \cite{23501,23502,6Gwhite}. This strategic integration also motivates the collaboration between well-known telecommunication operators (e.g. T-Mobile) and satellite network providers (e.g. SpaceX). Their goal is to provide worldwide communication services to users by establishing direct links between satellites and mobile phones \cite{t-mobile,optus}.

\begin{figure}[t!]
\centering
   \subfloat[]{\includegraphics[width=\linewidth]{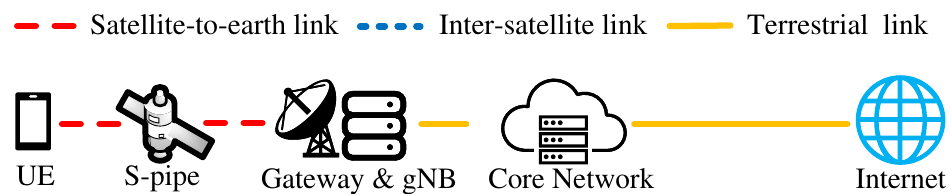}\label{pipe}}
   \quad
    \subfloat[]{\includegraphics[width=\linewidth]{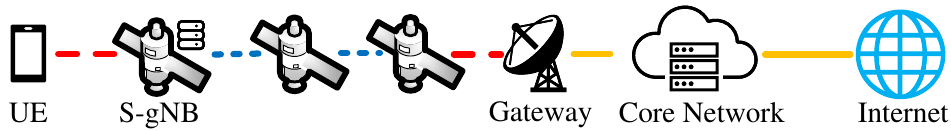}\label{regnerative}}
    \caption{Two different operating modes of mobile satellite network in (a) transparent mode and (b) regenerative mode.}
    
\end{figure}

There exist two operating modes in mobile satellite network, namely the transparent mode and the regenerative mode \cite{38821}. Initially, most satellites adopt the transparent mode, as shown in Fig.~\ref{pipe}. In this case, satellites serve as transparent physical nodes between user eqiupments (UEs) and ground stations. However, operating in this mode suffers from several obvious shortcomings, including limited coverage, single-point bottleneck, and relatively high latency \cite{stateless_mobile}. Recently, the regenerative mode, where LEO satellites act as base stations and provide worldwide coverage by leveraging inter-satellite links (ISLs), has been proposed and attracts extensive attention, as described in Fig. \ref{regnerative}. Meanwhile, mobile satellite networks operating in the regenerative mode have been tested \cite{re-test}. In the following discussion, we focus on the regenerative mode and refer to these satellites as S-gNB (\textbf{S}atellite next \textbf{G}eneration \textbf{N}ode\textbf{B}). 

During the development of mobile satellite networks, handover between satellites has emerged as a critical challenge. Different from the terrestrial network, the S-gNB, as a crucial infrastructure for user access located at LEO satellite, travels at a high speed and is usually far away from the core network. In the experiment on the mobile satellite network platform, ground terminals connected to the Starlink constellation experience handovers switched between two satellites every 2-5 minutes, resulting in an average handover latency of around 400 ms. This high-latency handover declines the user experience, especially for latency-sensitive applications.

In the earlier work, a prediction-based handover acceleration scheme was proposed which can considerably reduce the handover latency \cite{infocom24handover}. This scheme is based on the predictable traveling trajectory and unique spatial distribution of LEO satellites, to eliminate the interactions between the access and core networks during the handover process. However, this work is based on modifications to the Xn-based handover standard, while recent protocols \cite{23501} recommend conditional handover as a preferred solution for satellite networks. This change introduces new challenges for handover scheduling.

In this paper, we propose a parallel handover mechanism for mobile satellite networks, where handover occurs simultaneously in both the access and core networks rather than sequentially. The innovation behind the proposed scheme is a plan-based handover strategy instead of relying on accurate estimates of channel quality. Specifically, by introducing machine learning-based channel prediction, the RAN can effectively plan the timing of handovers and synchronize with the core network through a newly introduced network function. This approach allows to mitigate frequent interactions between the access and core networks, and thus eliminates the significant time overhead during the handover procedure.

The implementation of the proposed handover scheme entails two main challenges. First, to minimize the deployment cost, it is essential to ensure that the proposed solution is compatible with the existing 5G core network architecture. Secondly, it is crucial to ensure that the introduced handover scheme does not compromise the quality of service for users due to the unavailability of real-time measurement information. To address the first challenge, we introduce a novel network function which is referred to as Satellite Synchronized Function (SSF). In this way, it can facilitate seamless integration with the existing core network elements without modifications. Meanwhile, we propose a machine learning-based channel quality prediction model, which effectively improves the accuracy of channel prediction to address the second challenge.  Furthermore, by making use of the access strategy and spatial distribution characteristics in LEO satellite networks, it can alleviate the computational burden on the core network caused by numerous predictions.

Finally, we build a prototype which achieves the proposed handover scheme. This prototype mainly consists of modified UERANSIM and Open5GS and is driven by real LEO satellite traces including Starlink and Kuiper \cite{Kuiper}. Based on this prototype, we conduct extensive experiments. The results demonstrate that the proposed handover scheme can considerably reduce handover latency (around 21$\times$) and improve user-level performance such as TCP compared to three existing handover strategies.

Additionally, we also evaluate the performance of the prediction algorithm and the impact of user mobility. The results validate the feasibility of the proposed handover scheme in a large-scale mobile satellite network. 

The main contributions of this paper can be summarized as follows:

\begin{itemize}

\item To the best of our knowledge, this work represents the first research efforts to address the high latency problem of handover in mobile satellite network, which is an integral part of 5G and beyond NTN. Meanwhile, we have also demonstrated its deployment feasibility, since it brings no modification to the core network by introducing a novel component named Satellite Synchronized Function. 

\item We for the first time demonstrate how to decouple the interaction with the core network from the handover procedure by leveraging several intrinsic features in LEO satellite network such as predictable satellite trajectory and unique spatial distribution. 

\item We propose a novel hybrid signal strength prediction framework that integrates physics-based modeling with data-driven learning to enhance signal strength prediction accuracy. This dual approach ensures robust and precise predictions, even under dynamic conditions such as obstacles and weather variations. 

\item We have built an experimental prototype and conducted extensive experiments that are driven by real satellite traces including Starlink and Kuiper. Results verify that the proposed handover scheme can reduce the handover latency by around 21$\times$ compared to three existing handover strategies, thereby enhancing network stability and user-level performance, along with significant improvements in network stability and user-level performance. We are confident that our proposed approach serves as an effective advancement toward 6G NTN networks.

\end{itemize}
The rest of this paper is structured as follows. Section II introduces the background of the problem and our motivation. Section III gives an overview of our design. Section IV provides detailed explanations of the two aspects of our design. Section V describes our experimental setup and result. Section VI presents a review of related work in the field. Section VII discusses additional considerations and issues related to our work. Finally, Section VIII briefly concludes this work.

\section{Background} \label{chapect 2}

\subsection{LEO Satellite Nework}
In recent years, LEO satellite networks have drawn widespread attention due to the rapid advance in manufacturing and launch technologies. Based on the large-scale deployment of LEO satellites, UEs on the ground can enjoy low-latency and high-bandwidth network services with global coverage. Actually, LEO satellite networks can complement and integrate with traditional terrestrial networks to provide more robust and disaster-resistant communication services. According to recent reports \cite{InvestingStarlink}, millions of users have already benefited from satellite network services. 

Early satellite constellations, such as the initial versions of Starlink or Kuiper constellations, lacked the capability for direct inter-satellite data transmission. In this scenario, satellites act as relays between users and ground stations, necessitating that each satellite's view of the Earth includes at least one available ground station. Consequently, service providers must deploy and maintain a large number of ground stations worldwide to ensure service availability.

To address this issue, ISLs were introduced into LEO satellite networks. ISLs primarily utilize laser communication technology, enabling data transmission at exceptionally high rates. By establishing data links between adjacent satellites, the entire satellite constellation forms a connected network topology, significantly reducing the dependence on ground stations. In addition, ISLs help to decrease latency in data transmission across the network and improve overall bandwidth capacity.

Establishing laser links introduces certain hardware costs, and each satellite can only support a limited number of ISLs. Different satellites may have varying capacities to create ISLs depending on their specific design. Typically, a satellite can establish 3-5 ISLs with satellites in the same orbit or adjacent orbits \cite{9393372,starlink}. In this paper, we assume that each LEO satellite builds four ISLs with nearby satellites, consisting of two intra-orbit links connecting to the preceding and succeeding satellites in the same orbit, and two inter-orbit links connecting to satellites in adjacent orbits.

\subsection{Handover in Mobile Satellite Network} \label{handover in mobile}

\begin{figure}[t!]
	\centering
	\includegraphics[width=0.95\linewidth]{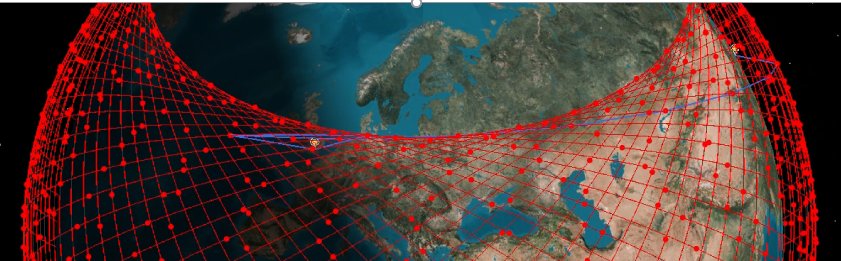}

	\caption{Illustration of handover signaling transmission (London to Shanghai).}

	\label{lts}
\end{figure}

Mobile satellite network follows the handover procedure specified in 5G standard \cite{38821}. There exist two kinds of handover schemes, respectively as Xn-based and N2-based schemes \cite{23501}. By leveraging ISLs, the Xn-based handover strategy can provide lower handover latency compared to the N2-based one. The high-level Xn-based handover procedure can be summarized as follows. First, UE disconnects the built Radio Resource Control (RRC) link with the source S-gNB and establishes a new RRC satellite-ground connection with the target S-gNB. Secondly, the target S-gNB delivers the handover results to the User Plane Function (UPF) through the ISLs and satellite-ground link, which is a critical part of the core network.  

As a measurement-enhanced handover scheme based on the Xn-based scheme, conditional handover (CHO) is considered by 3GPP as a baseline handover scheme for NTN \cite{38821}. In NTN, the signal strength of cells is very similar across the entire visible range of the satellite, so handover cannot rely solely on radio measurements. In the CHO, the source gNB establishes Xn connections with multiple potential target gNBs and notifies all potential target gNBs to pre-reserve the resources required for UE. The UE then decides on the target gNBs to access based on comprehensive information at the actual handover time, such as time and location information. 
The core idea of this scheme is to enhance the robustness of handovers in NTN scenarios by pre-establishing multiple Xn connections and pre-allocating resources, thereby avoiding the potential issue of signaling loss due to unstable satellite-ground connections during handover.
However, during the execution of CHO, interaction between the gNB and the core network is still required to switch the downlink. Therefore, the CHO scheme cannot fundamentally resolve the issue of prolonged handover latency in mobile satellite networks.

According to the above framework, there exists a key challenge (i.e., the handover problem) in mobile satellite network. Typically, handover in mobile satellite network differs from the terrestrial network in two main aspects\textemdash which will be illustrated as below. 

\begin{table}[t!] 
	\centering  
	\caption{Orbit information for commercial constellations} 
	\label{Orbit}  
	\begin{tabular}{|c|c|c|c|c|}  
		\hline  
		& & & &  \\[-6pt] 
		&\makecell{Altitude \\ ($km$)}& \makecell{Min. \\ Elevation  ($^\circ$)}&\makecell{Speed \\ ($km/s$)}&\makecell{Avg. Coverage \\ Time ($s$)} \\  
		\hline
		& & & & \\[-6pt]  
		Starlink&550&40&7.8&132 \\
            \hline
		& & & &\\[-6pt]  
		Kuiper&630&35&7.5&187 \\
		\hline
	\end{tabular} \label{Oi}

\end{table}

First, ground UEs experience frequent handover due to the fast traveling speed and small coverage of LEO satellites (compared with GEO satellites). OOn average, a handover occurs every three minutes. Actually, the handover interval may be even shorter because of the weather and UE’s access strategy. Table \ref{Oi} depicts the orbit information for two well-known commercial constellations. From this table, we can see that LEO satellites operating at different altitudes have different speeds and thus different time duration for connection. For both of these two constellations, the handover interval is less than 3 minutes because of the limited coverage time for each satellite.

Secondly, another distinguishing feature from the terrestrial network is that the mobile satellite network can provide seamless coverage all over the world even in deserts or oceans, which implies that the distance between UEs and the core network can be much longer than the terrestrial network \cite{6Gwhite,Sateliot}. Consequently, ISLs should be utilized to achieve handover signaling delivery between the access and core networks, which brings much higher latency. For example, Fig. \ref{lts} describes the signaling delivery between Shanghai and London, the distance of which reaches up to 9,000 km. Obviously, this incurs a rather long transmission latency, as well as a large handover latency. Moreover, due to the limited number of ISLs that can be established for each satellite, circuitous transmission paths between the access and core networks are common with more ISLs involved, thereby prolonging the propagation and handover latency.

\section{Design Overview}

\begin{figure}[t!]
\centering
   \subfloat[Starlink]{\includegraphics[width=0.45\linewidth]{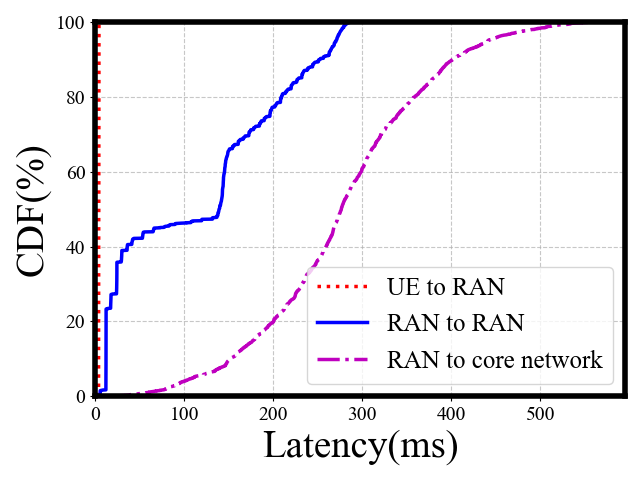}\label{static}}
    \subfloat[Kuiper]{\includegraphics[width=0.45\linewidth]{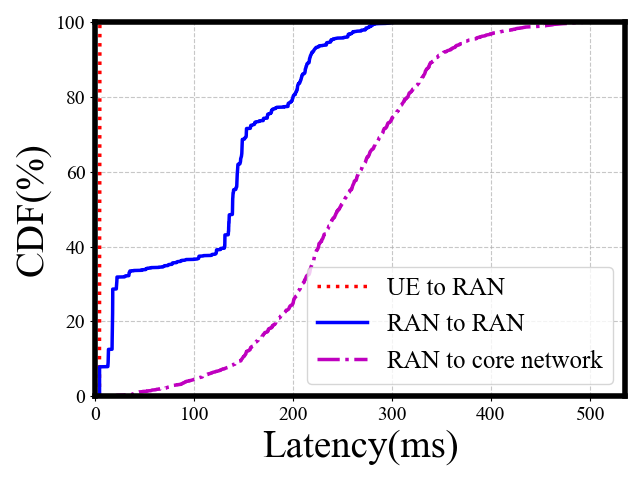}\label{static_Kuiper}}
    \caption{
  Transmission latency of different parts in handover.}\label{SSTAIC}
\end{figure}
The handover procedure proposed by the 3GPP 5G standard includes the switching of both the uplink and downlink paths. The uplink path switch is initially completed at the access network, followed by the transmission of signaling to the core network. Subsequently, the core network completes the switch of the downlink path. In this process, the downlink handover must occur later than the uplink handover, imposing a constraint on the overall handover speed. Furthermore, the transmission latency in the satellite scenario also plays a crucial role in influencing the latency of the handover. We conducted preliminary experiments in to investigate the transmission latency in handover. As shown in Fig.~\ref{SSTAIC}, we find that the time overhead consumed by the transmission from RAN to the core network dominates the overall transmission handover latency. The reason is that this process consists of passing through multiple ISLs and the satellite-ground link. Therefore, the main aim of this work is to parallelize the execution of the uplink path switch and downlink path switch in the handover procedure. This approach not only addresses the impacts introduced by sequential path switching but also avoids the transmission latency from the RAN to the core network, thereby significantly reducing the handover delay in mobile satellite networks.

\begin{figure}[t!]
	\centering
	\includegraphics[width=0.95\linewidth]{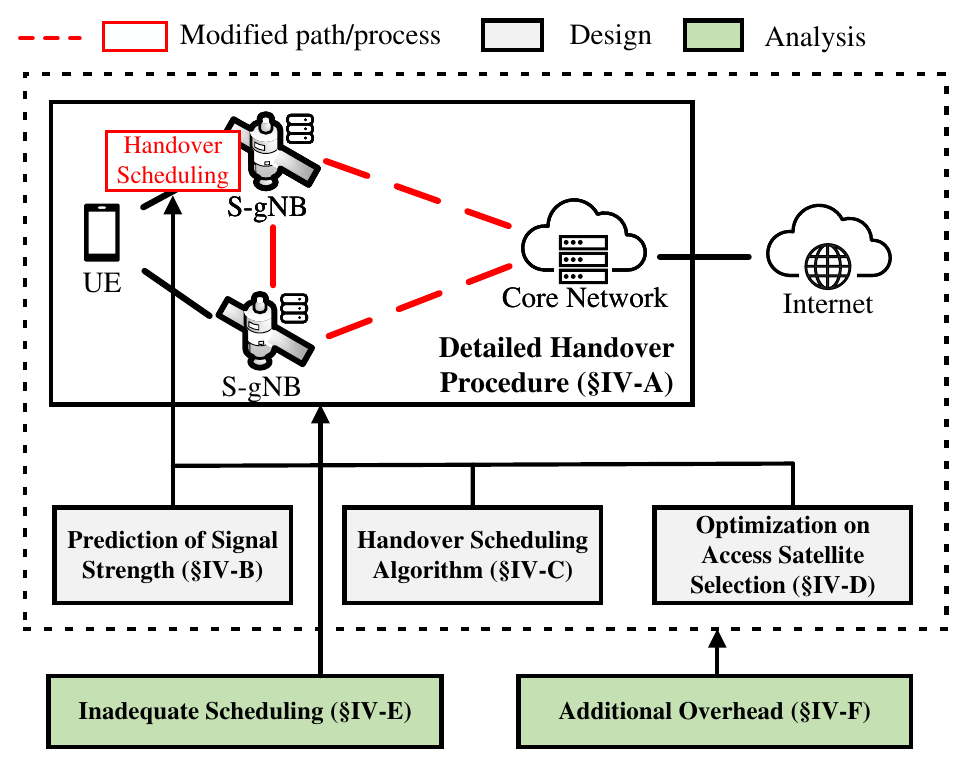}
	\caption{Overview of the Phandover design.}

	\label{overview}
\end{figure}

 In this paper, we design a concurrent handover scheme, {named Phandover, whose overall design is illustrated in Fig. \ref{overview}. In this scheme, the uplink and downlink path switches in the handover procedure are parallelized based on handover scheduling, thereby reducing the handover delay.}

However, this design introduces two new challenges. The first challenge is related to the compatibility between the proposed scheme and the existing 5G core network. In the proposed concurrent process, the access network no longer sends signaling during handover, which implies that the existing core network function may not support this procedure. The second challenge arises when real-time measurement reports from UEs are unavailable, potentially leading to unreasonable handover scheduling and a decline in user service quality.

To address the first challenge, we introduced a new network function, named as Satellite Synchronized Function (SSF), and designed a corresponding handover process (§\ref{detail_procedure}). Specifically, we divided the process into preparation and execution phases. In the preparation phase, SSF pre-stores the required signaling, and during the handover, it masquerades as the source S-gNB to send the stored signaling.

To address the second challenge, we propose a handover scheduling algorithm (§\ref{Synchronized}) that efficiently leverages the prediction of channel quality for scheduling handover targets and timing. To ensure the effectiveness of the algorithm, we also introduce a machine learning-based signal strength prediction model (§\ref{sec:signal}).

Furthermore, we have introduced several extra designs to reduce the handover latency and improve the system's robustness. For example, we add a simple constraint (i.e., selecting the access satellite in similar travelling direction with the previously connected satellite) in the satellite access scheme (§\ref{Access_selection}). Later analysis and experiments will demonstrate this easily achievable aim can considerably reduce the handover latency. Simultaneously, we investigate the impact of inaccurate prediction on the performance of the proposed handover scheme (§\ref{users_movement_fail}). To address its two main causes\textemdash user mobility and deviation in satellite trajectory prediction, we discuss and design corresponding strategies to deal with potential failures.

\section{Handover Design}

In this section, we provide a detailed explanation of our proposed handover scheme. First, we present a comprehensive overview of the proposed handover signaling procedure (§\ref{detail_procedure}). 
Then we delve into the two key designs: the signal strength prediction approach (§\ref{sec:signal}) and the handover plan algorithm (§\ref{Synchronized}). After that, we further optimize the access satellite selection scheme to reduce the handover latency (§\ref{Access_selection}). Subsequently, we introduce how we deal with the inaccurate prediction caused by user mobility and deviations in satellite trajectory prediction (§\ref{users_movement_fail}). Finally, we analyze the additional overhead introduced by the proposed method (§\ref{additional overhead}).
 
\subsection{Detailed Handover Procedure} \label{detail_procedure}

\begin{figure}[t!]
	\centering
	\includegraphics[width=\linewidth]{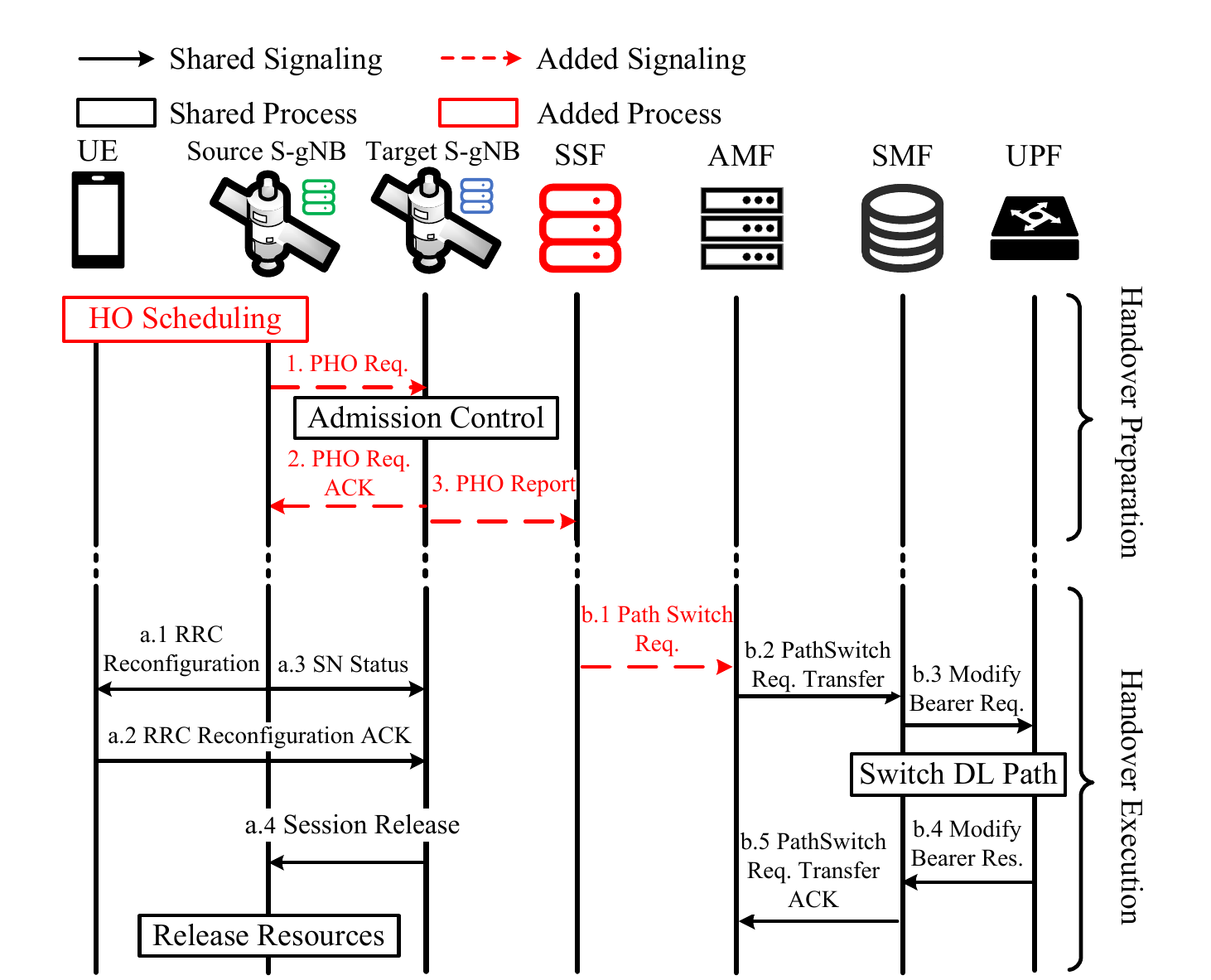}
	\caption{Comparison of standard and our proposed handover procedure in mobile satellite network.}
 
	\label{handover_procedure}
\end{figure}

As shown in Fig.~\ref{handover_procedure}, the proposed handover process can be divided into two main parts: handover preparation and handover execution. During the handover preparation phase, the RAN determines the timing and target for the handover, transmitting the required information to the core network. In the handover execution phase, both the RAN and the core network synchronously execute the handover. The proposed handover procedure can be summarized as below.

First is the handover preparation phase: This phase occurs in the early stages of handover, during which the S-gNB determines the handover direction and communicates the result to the core network.\\
\textbf{Step 1} Based on calculations such as satellite trajectory predictions, the source S-gNB pre-determines the handover target S-gNB and timing. Subsequently, it informs the target S-gNB of the decision. \\
\textbf{Step 2} The target S-gNB prepares for the handover, such as pre-allocating channel resources. Subsequently, the target S-gNB confirms the pre-handover with the source S-gNB. \\
\textbf{Step 3} The results of this pre-handover are transmitted to the SSF, including all necessary information for the handover execution such as the UE's NGAP ID in the AMF and the target S-gNB, along with essential information to trigger the path switch, including the path switch timing.

After that is the handover execution phase. Based on the handover timing determined in the handover preparation phase, the core network and access network synchronize for the handover execution. \\
\textbf{Steps for the access network} Initially, there are two simultaneous operations. On one hand, the source S-gNB informs the UE of the handover decision. Subsequently, the UE disconnects from the source S-gNB and establishes a new RRC connection with the target S-gNB. These actions correspond to Steps a.1 and a.2 in Fig.~\ref{handover_procedure}. On the other hand, the source S-gNB transmits relevant synchronized information to the target S-gNB. This information includes the data to be transmitted and the sequence number (SN) (Step a.3).
At last, the source S-gNB is notified to release the resources (Step a.4) \\
\textbf{Steps for the core network} Firstly, the synchronization algorithm in Step b.1 propels the handover process, with the SSF initiating a Path Switch request. Following this, a standard path switch procedure is triggered, involving the UPF to facilitate the handover of the downlink path (Steps b.2 to b.5).

Here, we will elaborate on how the proposed procedure ensures compatibility with the existing core network. Fundamentally, for the core network, the SSF disguises itself as the S-gNB in the handover process\textemdash it sends signaling to the core network based on the gNB's handover plan. The implementation of handover preparation also shares some aspects to ensure compatibility. Overall, the proposed handover preparation aligns with the corresponding steps in the 5G handover. The only difference lies in the target S-gNB, where it pre-assigns the NGAP ID for the UE upon receiving a PHO request, while in the 5G standard approach, this assignment takes place after the RRC Reconfiguration completion. This modification aims to ensure that the SSF has all the necessary information before the handover execution.
 
Based on the above description, the key innovation in our scheme is the handover scheduling algorithm, which ensures that the network can execute the handover at the appropriate time without relying on real-time base station measurement reports. However, determining a reasonable handover time in advance poses a non-trivial challenge. For instance, inaccurate signal strength estimates can lead to the selection of inappropriate satellites for handover. On the other hand, a limited prediction frequency may result in an unreasonable timing choice for the handover. Additionally, the prediction operation for a large number of ground UEs imposes a significant computing burden on the S-gNB.

In the following subsections, we will provide a detailed description of the handover prediction algorithm, as well as how to mitigate the computation overhead, by leveraging two inherent features in LEO satellite networks\textemdash predictable trajectory and unique spatial distribution.

\subsection{Prediction of Signal Strength} \label{sec:signal}

\begin{figure}[]
	\centering
	\includegraphics[width=0.65\linewidth]{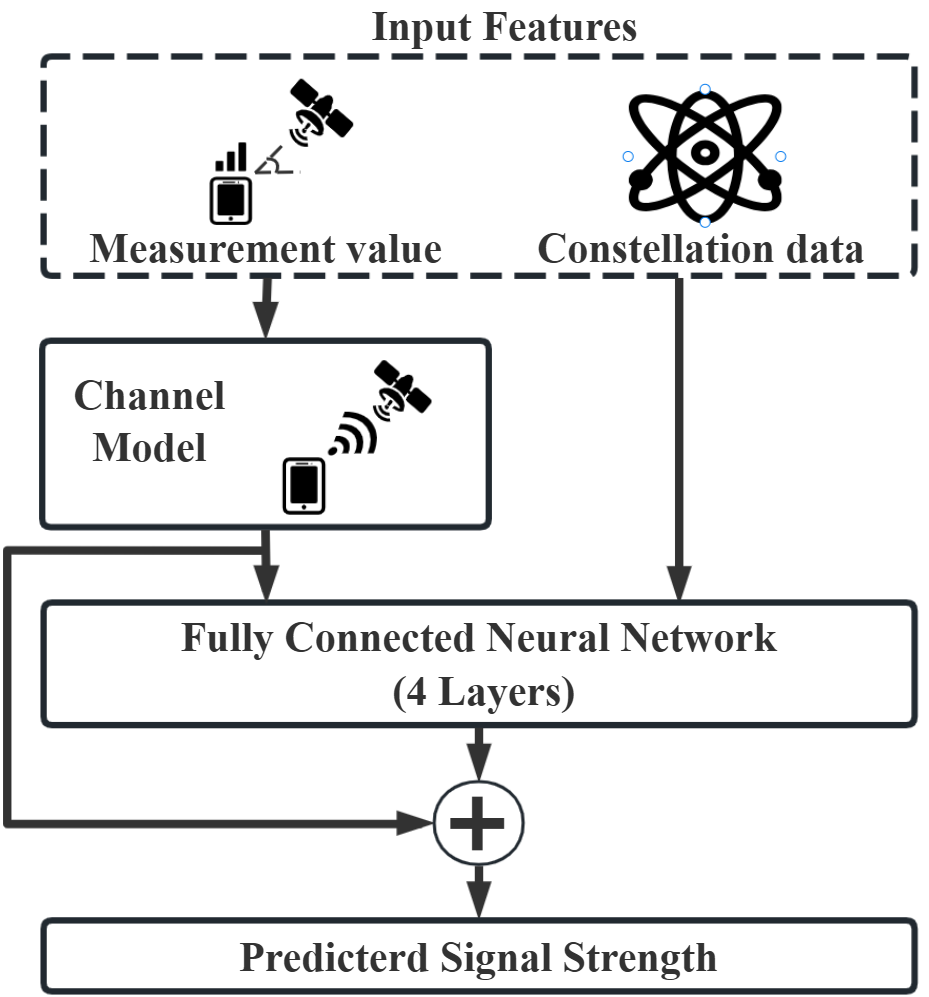}
	\caption{Architecture of the proposed machine learning model for signal strength prediction.}
 
	\label{model}
\end{figure}

In the mobile network, handovers are prompted by real-time UE channel measurement data conducted by the gNB. The gNB will make the handover decision when UE reports indicate that the signal strength of another gNB exceeds a certain threshold. Subsequently, the target gNB informs the core network of the handover outcome, and the core network executes the corresponding handover procedure. In the proposed parallel handover scheme, handovers between the core network and the RAN occur simultaneously. This necessitates the network to predict the occurrence of handovers, which requires predicting the signal strength of the satellite. This is crucial because inaccurate link quality predictions may result in users switching to satellites with poor link quality, thereby resulting in low transmission rates or high packet loss. 

The signal strength is influenced by various factors at multiple levels, including propagation distance, weather conditions, environmental factors, and antenna design. In consideration of these factors, the signal strength can be calculated using existing satellite-to-ground channel models. However, environmental elements such as trees and buildings surrounding UEs can introduce obstacles or reflections. The impact of weather on the signal is also challenging to accurately estimate. These factors contribute to empirically inaccurate results. 

To address the challenge of accurately predicting signal strength, we propose a novel hybrid prediction framework that integrates physics-based modeling with data-driven learning. Unlike purely empirical or theoretical approaches, our model leverages satellite-to-ground channel models to establish an initial estimate of signal strength while employing neural networks to adaptively learn environment-related features from historical data. This dual approach enables more accurate and robust predictions, even in the presence of dynamic environmental conditions such as obstacles and weather variations. Fig.\ref{model} illustrates the proposed prediction model, and the details of each module are described as follows.

\noindent\textbf{Input Features:} We formulate the problem as a multivariate-to-univariate prediction task. The input features of the model are denoted as \( [\theta, h, \theta_{0}, s_0] \),  $\theta$ is the elevation angle between the UE and the satellite, $h$ is the height of the satellite, $\theta_0$ and $s_0$ refer to an earlier measurement's elevation angle and signal strength. Our goal is to predict the signal strength at a specific time $t$, denoted as $s_t$. Notably, the model's input does not directly include historical data, as we utilize historical information to construct the training set, allowing the model to learn relevant patterns from it.

\noindent\textbf{Channel Model}: 

The signaling through the wireless channel will suffer propagation losses, leading to variations in the received signal strength. 3GPP  provides the path loss calculation formula as follows \cite{38811}:

\begin{equation}
PL = PL_b + PL_g + PL_s + PL_e \label{eq:pl}
\end{equation} 
where $PL$ is the total path loss, $PL_b$, $PL_g$, $PL_s$, and $PL_e$ are the basic path loss, attenuation due to atmospheric gases, attenuation due to either ionospheric or tropospheric scintillation, and building entry loss, respectively. All of these losses are measured in dB.

$PL_b$ is composed of Free Space Path Loss (FSPL), Shadow Fading (SF), and Clutter Loss (CL). SF is represented by a random number generated by a normal distribution, and CL is negligible under the scenario considered. $PL_g$ can be considered a loss proportional to the distance the signal travels through the atmosphere. but $PL_s$ is difficult to predict due to its rapid variation. $PL_e$ can be neglected in the scenario considered. 
These losses can be formulated as follows:

\begin{equation}
PL_b = FSPL(d, f) + SF + CL(\alpha, f) 
\end{equation} 
\begin{equation}
FSPL(d, f) = 32.45 + 20 \log_{10}(f) + 20 \log_{10}(d)
\end{equation}
\begin{equation}
SF \sim N(0,\sigma_{SF}^2)
\end{equation}
\begin{equation}
PL_g = \frac{A_{zenith}(f)}{sin(\alpha)} \label{eq:pl_last}
\end{equation}
where $d$ is the distance in meters, and $f$ is the carrier frequency in megahertz (MHz). $\sigma_{SF}$ refers to the standard deviation of shadow fading. $A_{\text{zenith}}$ refers to the corresponding zenith attenuation. 

Based on Eq. \ref{eq:pl} to \ref{eq:pl_last}, we can calculate the signal strength in time $t$ using the signal model, denoted as $s^{\text{model}}_t$, with the measured signal strength $s_{0}$, the position of the satellite and environmental parameters. We have:
\begin{equation}
s^{\text{model}}_t = s_{0} -20 \log_{10}(\frac{d}{d_0}) - \frac{A_{zenith}(f)}{sin(\alpha)} + \frac{A_{zenith}(f)}{sin(\alpha_{0})} \label{eq:model}
\end{equation}

\noindent\textbf{Hidden Layer}: The input data processed by the channel model is combined with the original input values and fed into the hidden layer. Considering the dominant role of the prediction model in the overall process, we introduce shortcut connections in the network, where \( s^{\text{model}} \) is added to the output of the fully connected layer. This design encourages the network to learn from the channel model, improving both the convergence speed and the accuracy of the network.

\subsection{Handover Scheduling Algorithm} \label{Synchronized}
\noindent \textbf{Main Procedure:}
From a high-level perspective, the proposed handover scheduling involves using predicted signal strength instead of actual measured signal strength in predicting the UE's access to S-gNB. In other words, the objective of the handover scheduling algorithm can be summarized as follows: for each UE, the S-gNB predicts its signal strength concerning itself and other satellites. Then the algorithm identifies a handover timing according to the handover strategy. For example, the handover strategy could involve switching to another satellite if its signal strength exceeds a certain threshold or if the channel quality of the current satellite falls below a specified threshold while another satellite offers sufficient quality. This strategy is determined by the satellite network's operator.
Once the scheduling for the UE handover time is completed, S-gNB proceeds with the subsequent operations of the handover process, as described in §\ref{detail_procedure}.

To achieve this goal, two main steps need to be implemented: first, the prediction of signal strength to anticipate the UE's access satellite at a given moment, and second, the development of a strategy for this prediction. In the context of the first step, we can calculate the relative position of the satellite to the user based on the satellite's orbital trajectory and then utilize the signal strength prediction method discussed in the previous section. As for the second step, predicting the access S-gNB at fixed time intervals $\Delta t$ is a straightforward approach. However, this simple method may cause the decision of the handover time to deviate from the optimal time by up to $\Delta t$. Since $\Delta t$ is typically in the order of hundreds of milliseconds due to the computing complexity of prediction, the UE may experience suboptimal service quality within that time frame.

Alternatively, we investigate the case of predicting the users' access satellites at two-time points simultaneously. The detailed description is shown as follows. Assume that $t_0$ and $t_1 = t_0 + \Delta t$ are two consecutive time points at which the S-gNB performs predictions. We use $U$ to refer to all UEs served by the S-gNB, and use $\mathcal{A}_t$ to represent the set of access satellites for all users $U$ at time $t$. By reasonably selecting $\Delta t$, we ensure that at most one handover is triggered for each UE $u \in U$ between $t_0$ and $t_1$. As a result, by comparing $\mathcal{A}_{t_0}[u]$ and $\mathcal{A}_{t_1}[u]$\textemdash the access satellites of user $u$ at time $t_0$ and $t_1$, we can determine whether a handover will be triggered during $t_0$ to $t_1$. 
If a handover is set to be triggered, the S-gNB needs to decide the handover timing to maximize the UE's service quality. To accomplish this, a simple yet effective binary search method can be employed to pinpoint the optimal time point. 

Consequently, we propose a handover scheduling algorithm, as shown in Algorithm \ref{Algorithm Synchronized}. Here, we use $\mathcal{A}_t$ to represent the optimal set of access satellites for all users at time $t$, stored in table $\mathcal{R}$. To distinguish it from real-time \( t \), we use \( T \) to denote the time when the S-gNB has completed handover scheduling. Specifically, by the time the real world reaches \( T \), the handover scheduling for \( T + \Delta t \) should have already been completed. 
The algorithm plans the switch for the next $\Delta t$ time frame with a periodicity of $\Delta t$. Let's take an example of a prediction for the time frame $T + \Delta t$ to $T + 2\Delta t$. First, we obtain the access satellites of all UEs at $T + \Delta t$—denoted as $\mathcal{A}_{T+\Delta t}$—from $\mathcal{R}$, and then predict the access satellites at $T + 2\Delta t$—denoted as $\mathcal{A}_{T+2\Delta t}$—according to the predicted trajectory of LEO satellites. Based on the difference between $\mathcal{A}_{T+\Delta t}$ and $\mathcal{A}_{T+2\Delta t}$, we employ a binary search algorithm to calculate the set of predicted handover triggering times $\mathcal{T}p$. Specifically, for each UE that will handover, we compute their associated access satellites at the intermediate time instant, i.e., $T+1.5 \Delta t$, to halve the error for handover timing. This process is iteratively repeated until the error becomes sufficiently small. Once we determine the handover times for all users to switch, these users proceed with the subsequent steps of the handover preparation process.

\begin{algorithm}[t]
\caption{Handover Scheduling Algorithm\label{Algorithm Synchronized}}   
{{
    \begin{algorithmic}[1]

    \State Initialize $T$, $\mathcal{R}$ 
    \While{current time $>T$}
    \State \Call{Handover Plan}{$T+\Delta t,T+2\Delta t$} 
    \Statex \Comment{Conduct predictions in advance by $\Delta t$.}
    \State $T = T +\Delta t $
    \EndWhile
    \Procedure{Handover Plan} {$t_{begin}$,$t_{end}$} 
    \State Get $\mathcal{A}_{t_{begin}}$ from $\mathcal{R}$
    \State Based on $\mathcal{A}_{t_{begin}}$, predict $\mathcal{A}_{t_{end}}$  
    \State According to $\mathcal{A}_{t_{begin}}$ and $\mathcal{A}_{t_{end}}$, get $\mathcal{T}_p$ with binary search 
    \State Execution handover preparation  
    \EndProcedure
    
    \end{algorithmic}
    }
    }
\end{algorithm}

\noindent \textbf{Fast Satellite Selection:} 
Given the limited computational capacity of satellites, the computational burden imposed by the handover scheduling algorithm on satellites is also a topic worth discussing. The S-gNB needs to select suitable access satellites for all UEs it serves every $\Delta t$. A direct prediction approach that considers all satellites in the constellation for every UE to predict the access satellite would impose significant computational pressure and make it difficult to complete within the specified $\Delta t$.

To address this challenge, we propose a fast access satellite selection algorithm aiming to reduce the number of predictions from both the user and satellite perspectives. First, we take into account the access strategies of UEs to reduce the number of users that need to be considered. We categorize the existing access strategies into two types: \textbf{consistent} and \textbf{flexible} \cite{4062836,Satellite_handover}. The former maintains a connection until the user leaves the service coverage area, while the latter may handover even when UE is within its access satellite's coverage area. For UEs with consistent access strategies, the algorithm simply checks if the S-gNB can still provide service, significantly reducing the number of UEs required consideration. 

Secondly, we further reduce the computation complexity by considering the distribution of satellites based on their positions. Instead of computing the situation for a UE with all satellites, we only need to consider those satellites that could be selected as handover targets. Specifically, we divide the Earth into several continuous rectangular blocks based on the service radius of the satellites and assign satellites to corresponding blocks according to their positions. As a result, for each S-gNB, only nearby satellites require computation, significantly reducing computation complexity. 

The specific steps of the algorithm are illustrated as Algorithm \ref{Algorithm 1}. Specifically, the algorithm can be divided into 3 steps as follows: \\
\textbf{Step 1}: We construct the UE candidate set \(U_C\). If the S-gNB selects a flexible strategy, we add only those whose access satellite is not available at \(t\) to the set \(U_C\) for further processing. And if a consistent strategy is selected, we add all UEs to the set \(U_C\). \\
\textbf{Step 2}: Utilizing orbit information, we construct a candidate set \(\mathcal{SAT}_C\) consisting of satellites within a certain range of the current satellite.\\
\textbf{Step 3}: For each user \(u_j\) in \(U_C\), based on signal strength information, we find the corresponding access satellite \(sat_j\) in \(\mathcal{SAT}_C\), i.e., \(\mathcal{A}_{t}(u_i) = {sat}_j\).

\begin{algorithm}[t]
\caption{Fast access satellite selection algorithm\label{Algorithm 1}}   
{{
    \begin{algorithmic}[1]
    \Require $U$, $t$
    \Ensure $\mathcal{A}_{t}$
    \State Initialize UE candidate set $U_C$, Satellite candidate set $SAT_C$, $A_{t}$

    \ForAll{$u_i \in U$}
    \If{Access strategy is consistent} 
    \State Predict signal strength between $u_i$ and the S-gNB $sat$
        \If{the satellite is not available for $u_i$ at {t}}
            \State  $U_C = U_C \cup \{u_i\}$
        \Else
            \State $\mathcal{A}_{t}(u_i) = \mathcal{A}_t(u_i)$
        \EndIf
    \ElsIf{Access strategy is flexible}
        \State  $U_C = U_C \cup \{u_i\}$
    \EndIf
    \EndFor

    \State Put all satellites near to $sat$ into ${SAT}_C$
    
    \ForAll{$u_j \in U_C$}
    \State From $SAT_C$, find the access satellite ${sat}_j$ of $u_j$ 
    \State $\mathcal{A}_{t}(u_i) = {sat}_j$ 
    \EndFor
    \State \Return $\mathcal{A}_{t}$
    \end{algorithmic}
    }

    }

\end{algorithm}

\noindent \textbf{Parameter Selection:} The remaining issue is to explore the appropriate value set for the update interval $\Delta t$, which should ensure that two consecutive handovers for any stationary UE do not occur within $\Delta t$. A feasible solution is to set $\Delta t$ to be less than the constellation's minimum service time, which is influenced both by the satellite constellation configuration and the perhaps additional limitation when selecting access satellites (to avoid excessive frequent handovers). Based on the analysis of existing satellite constellation configurations and actual measurements \cite{neinavaie2022unveiling,starlink,Kuiper,oneweb}, we set the update interval $\Delta t$ to 5 seconds. Consequently, by iterating the binary search process 9 times, we can attain a prediction accuracy as precise as 10 milliseconds, meeting the specified requirements.

\subsection{Optimization on Access Satellite Selection} \label{Access_selection}
To further reduce the overall handover latency, we focus on minimizing the latency between source S-gNB and target S-gNB, which is the time taken for the handover process. Based on the preliminary experiments shown in Fig. 3, we can observe a sudden increase in latency in 30-40 ms. The main reason is that the satellite switches in contrary directions (i.e., handover from a northern-direction satellite to a southern-direction satellite, or vice versa) \cite{chen2021analysis,zhang2022enabling}.
To this end, we make an additional constraint that handovers can happen only between satellites in a similar-direction, which is expected to reduce the propagation delay between satellites by at most more than 200ms.

\subsection{Inadequate Scheduling} \label{users_movement_fail}
\begin{figure}[t!]
	\centering
	\includegraphics[width=\linewidth]{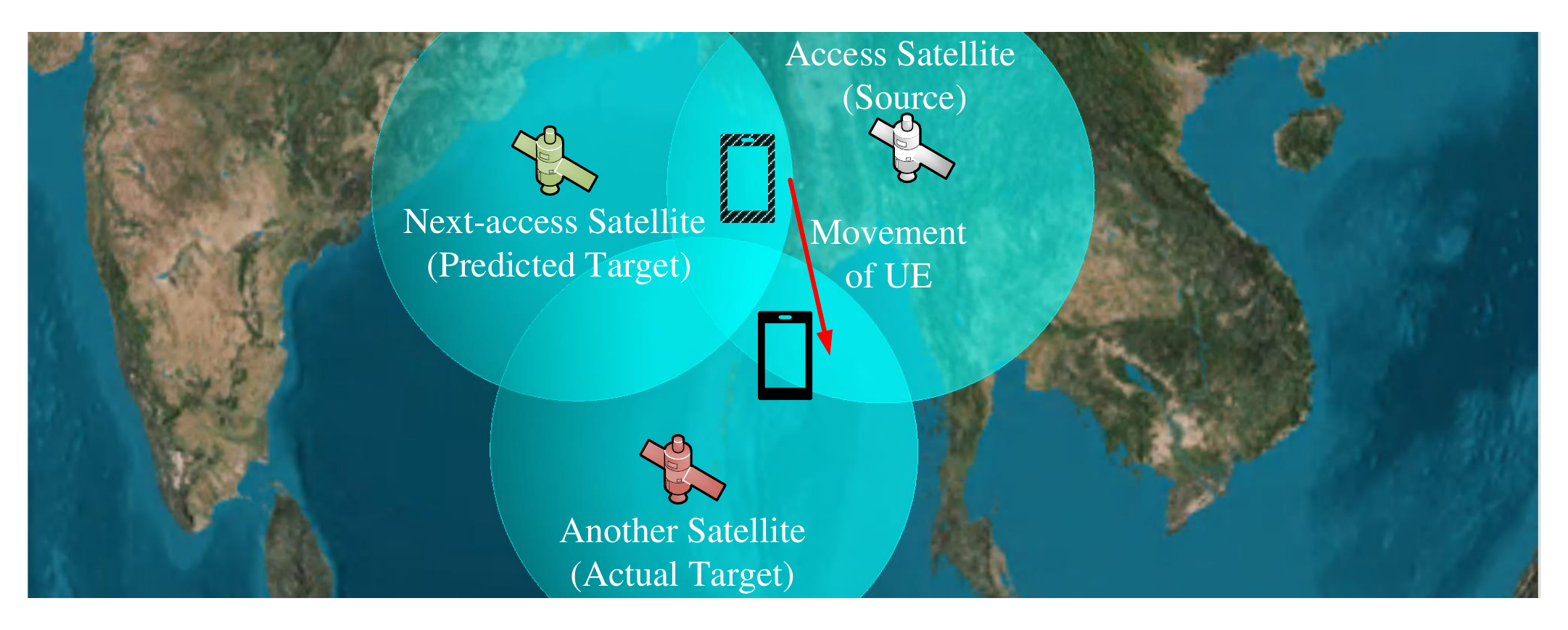}
	\caption{Illustration of inadequate handover scheduling caused by user mobility.}
	\label{uemobile}
\end{figure}
Beyond the predictive errors in signal strength, several factors may contribute to the formulation of ‘inadequate handover scheduling', where the planned handover timing or target diverges from the original measurement-based handover. Such disparities can result in diminished service quality. We posit that user mobility and deviations in satellite trajectory are pivotal factors contributing to these discrepancies. In the ensuing discussion, we will elaborate on the impact of these two factors and propose solutions.

On one hand, in the context of users, despite their significantly lower speed compared to LEO satellites, there exist some cases where user mobility results in inadequate handover scheduling. Such cases may occur when the user move at high speeds and is located at the overlapping region of the satellites' service coverage, as illustrated in Fig. \ref{uemobile}. 

It is crucial to highlight that these scenarios are infrequent, even for UEs who are moving at high speeds. Consequently, it can be regarded that user mobility has little impact on the user experience during the handover process. In \ref{results of Experiment}, we will delve into the probability of inadequate handover scheduling using experimental data.

On the other hand, deviations in satellite trajectory prediction can also lead to inadequate handover scheduling. The primary cause of these deviations is the accumulation of trajectory prediction errors. Given the complexity of perturbation, commonly used satellite orbit models, such as SGP4, introduce errors that gradually accumulate as the satellite moves. As a result, trajectory prediction based on daily-updated ephemeris data may exhibit errors of several kilometers \cite{orbitprediction}, ultimately leading to abnormal handovers with a probability of approximately $10^{-3}$.

To tackle this problem, we employ short-term trajectory prediction based on minute-level updated satellite ephemeris data, whose prediction precision reaches 10 centimeters \cite{rs15010133}; thus, avoiding the most inaccurate predictions. 

\subsection{Additional Overhead}
\label{additional overhead}

The additional overhead of the proposed scheme primarily stems from periodic handover scheduling and the forwarding and storage of handover commands, which introduce computational, communication and storage overhead.

\noindent\textbf{Computational Overhead:} In the proposed scheme, at every \(\Delta t\) interval, the serving satellite of each user must perform a handover scheduling. During each scheduling process, in the worst-case scenario, the serving satellite needs to evaluate the signal strength of all potential target satellites within a certain range. Consequently, the total number of comparisons per unit time for the entire system can be expressed as:  

\begin{equation}
    C_{\text{cal}} =  \frac{M * N *\phi}{\Delta t}
\end{equation}

where \( M \) is the total number of ground users served by the system, \( N \) is the total number of satellites. \( \phi \) represents the ratio of the number of satellites considered for user handover to the total number of satellites, which is related to the satellite's coverage area and the constellation configuration. 

\noindent\textbf{Communication and Storage Overhead:} In the proposed scheme, the SSF stores users' handover signaling, i.e., \textit{PathSwitchRequest}, and transmits them at the appropriate time. This mechanism introduces additional communication and storage overhead. Specifically, the total additional communication overhead in the system can be expressed as:  

\begin{equation}
    C_{\text{comm}} = M * f * S
\end{equation}
and the additional storage overhead can be represented as:
\begin{equation}
    C_{\text{storage}} = M * f * S * \Delta t
\end{equation}
 where \( M \) is the total number of users, \( f \) is the handover frequency per user, \( S \) is the average size of a handover signaling message 

Below, we provide an estimate of the communication and storage overhead. Consider a scenario in which the total number of users served is $M=300,000$, with a Starlink satellite constellation and a single signaling message size of approximately 500B. The proposed scheme would introduce an average additional network overhead of about 1MB per second on the core network and an additional storage overhead of approximately 5MB, which remains entirely acceptable for practical deployment.

\section{Experiments}

\subsection{Experimental Setup}

\begin{figure}[t!]
\centering
   \subfloat[]{\includegraphics[width=0.23\textwidth]{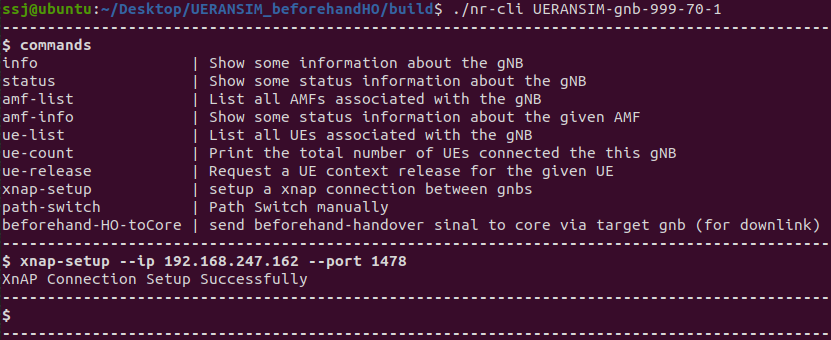}\label{Xnap_Establish_fig}}
   \hspace{0.1in}
    \subfloat[]{\includegraphics[width=0.23\textwidth]{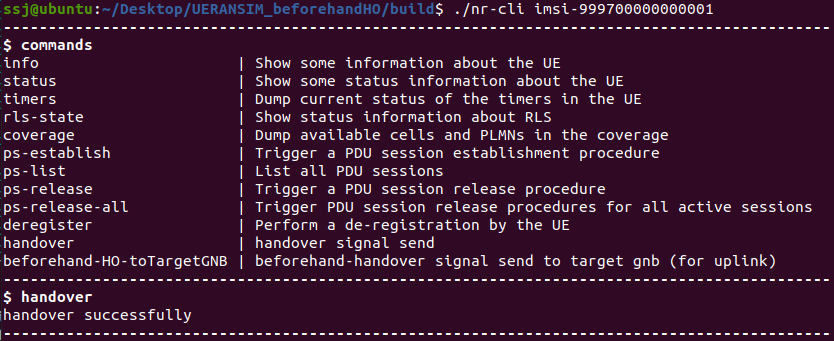}\label{handover_success}}
    \caption{
  Developed Xn-based handover: (a) Xnap connection buildup; (b) Handover success.}\label{Xn-based}
\end{figure}

\noindent \textbf{Satellite Constellation}: Experiments are carried out driven by the Starlink and Kuiper constellation traces that are obtained from \cite{celestrak}. And we can also get the information of ground stations from \cite{StarlinkStatus}. 
Finally, we have built a platform for simulating real-trace dynamics of LEO satellite constellation using skyfield \cite{skyfield} based on these constellation and ground station data.

\noindent \textbf{Signal Strength}:
To validate the effectiveness of the channel prediction algorithm, we utilized real-world signal strength data from four different LEO satellites, namely NOAA-20/JPSS1, AQUA, TERRA, SNPP, and two ground stations located at different positions: one in GUAM (GM) and the other in Wisconsin (WI) \cite{L2D2}. We treated the ground stations as UE and conducted experiments for each of the eight satellite-UE pairs. Each experiment involved 30 days of data collection for the respective pairs.

\noindent \textbf{System Prototype}: Driven by the above platform, we have built a prototype by combining UERANSIM \cite{ueransim} with Open5GS \cite{open5gs} for 5G and beyond non-terrestrial network, which achieves handover in mobile satellite networks. UERANSIM is a widely utilized simulator for both UE and S-gNBs implementation in 5G network, while Open5GS is an open-source implementation of the 5G core network. Following the 5G standard signaling flow, we have modified UERANSIM to support the Xn-based handover, as shown in Fig. \ref{Xn-based}. The built prototype operates on a commodity laptop with a 2.5 GHz CPU core and 16 GB RAM. Additionally, we also deployed the machine learning model for signal strength prediction on the laptop.  

\noindent \textbf{Comparison handover schemes}: We compare our proposed handover scheme with the following three handover schemes, in order to demonstrate the higher efficiency of the proposed handover strategy.
\begin{itemize}
\item \textbf{NTN} refers to the standard handover process specified in 5G NTN, which is described in §\ref{handover in mobile}. We compare the proposed handover scheme with the NTN handover scheme to show the performance improvement brought by the modified handover. 

\item \textbf{NTN-GS} refers to the handover procedure assisted by nearby ground stations, which implies that ground stations close to the LEO satellite are leveraged to record handover information, inspired by the handover strategies designed in IP satellite network \cite{dong2018mianxiang,7811041}. Thus, the controlling signaling for handover does not need to be transmitted to the core network and then the handover latency can be reduced.

\item \textbf{NTN-SMN} refers to the handover procedure assisted by nearby space network (SMN) \cite{9755268}. Similar to the NTN-GS handover scheme, satellites function as both the access and core network simultaneously in this strategy. This approach helps reduce handover latency as well.

\end{itemize}

\noindent \textbf{Signal Strength Prediction}:  
To evaluate the precision of the proposed handover approach, we compare it with the following two signal strength prediction schemes.  

\begin{itemize}  
\item \textbf{ITU} refers to a signal strength prediction approach based on the channel model provided by 3GPP \cite{38811}. Specifically, it utilizes the satellite's position and estimates the signal strength based on Eq. \ref{eq:model}.  

\item \textbf{LSTM} refers to a signal strength prediction approach based on the LSTM \cite{lstm} algorithm, a classic machine learning algorithm for time-series prediction. In this approach, the past 10 seconds of data are used to predict the signal strength at the next second. Consequently, the input parameter size for this method is ten times larger than that of the proposed model. However, LSTM can only predict signal strength at fixed time points, making it unsuitable for the proposed approach. Therefore, it serves as a baseline for accuracy comparison.
 
\end{itemize}

\subsection{Experimental Results} \label{results of Experiment}

\begin{figure*}[t]
\centering
\subfloat[]{\includegraphics[width=0.333\textwidth]{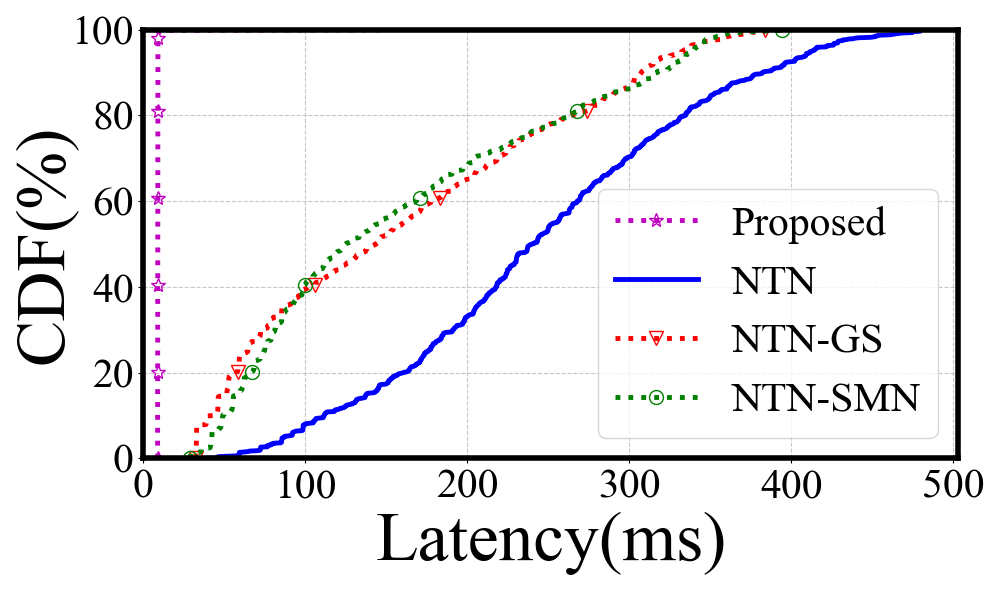}\label{latency}}
\subfloat[]{\includegraphics[width=0.333\textwidth]{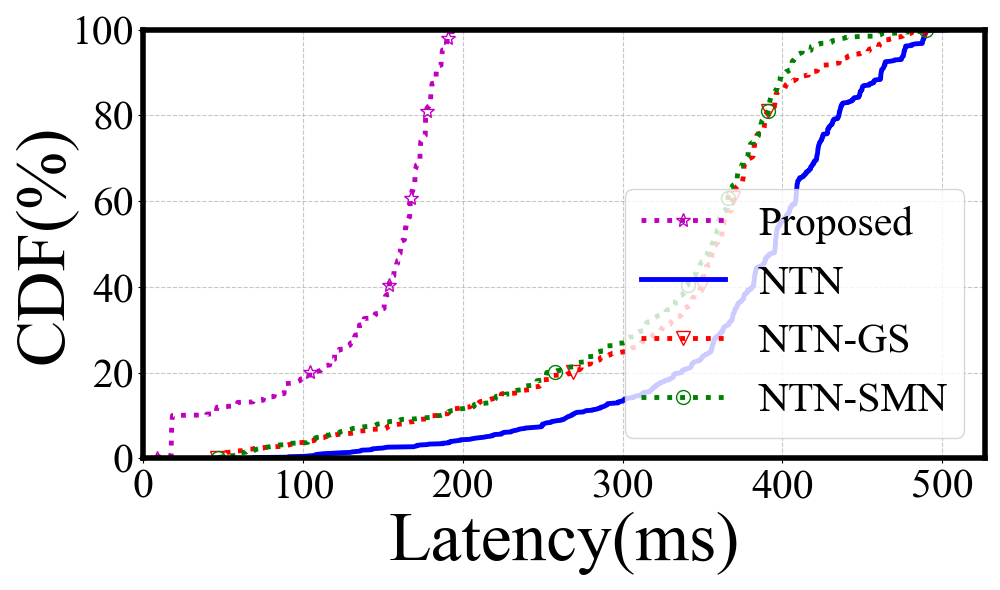}\label{latency_all_direction}}
\subfloat[]{\includegraphics[width=0.333\textwidth]{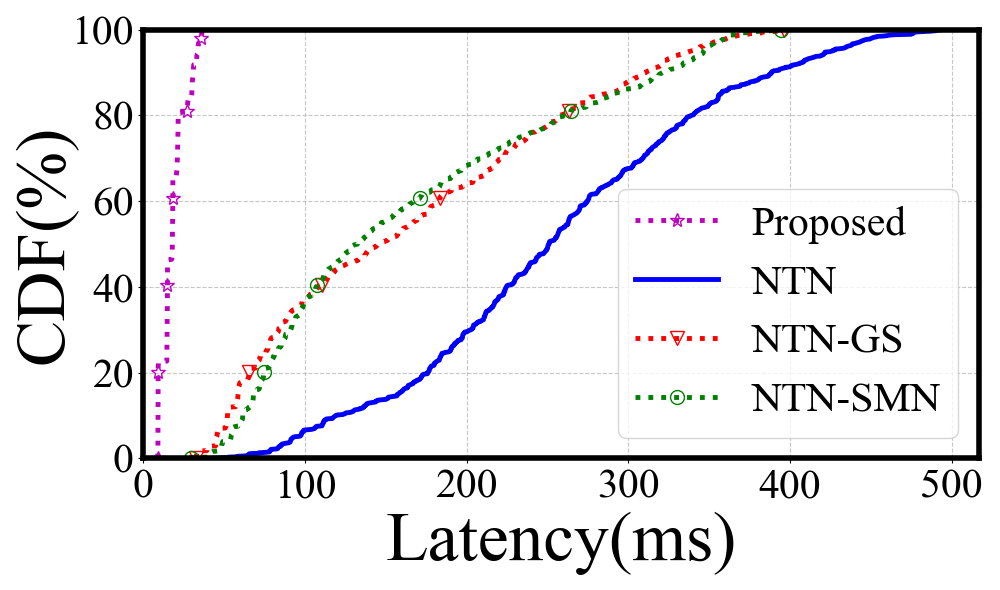}\label{latency_consistent}}
\quad
\subfloat[]{\includegraphics[width=0.333\textwidth]{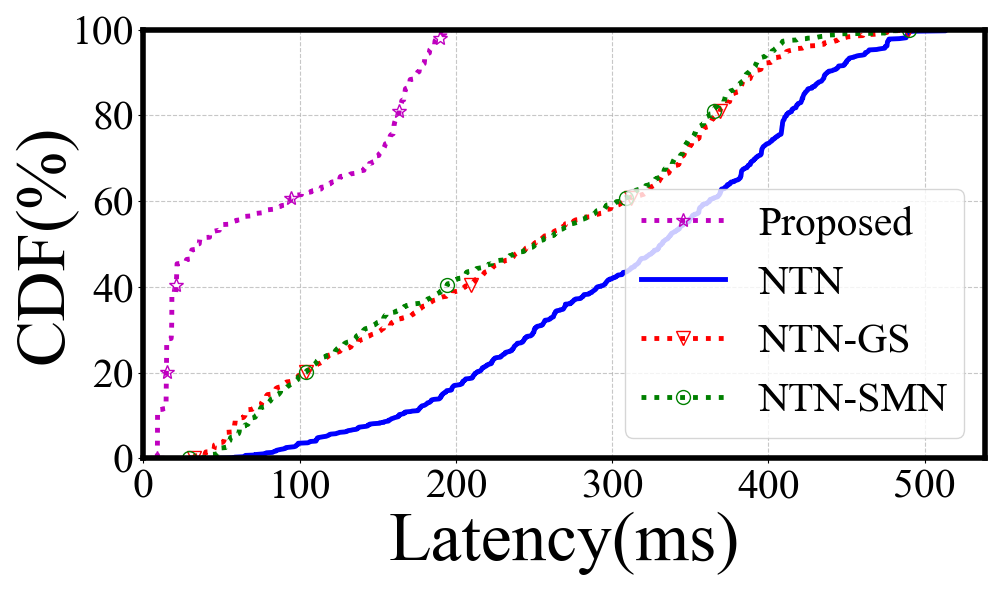}\label{latency_consistent_all_direction}}
\subfloat[]{\includegraphics[width=0.333\textwidth]{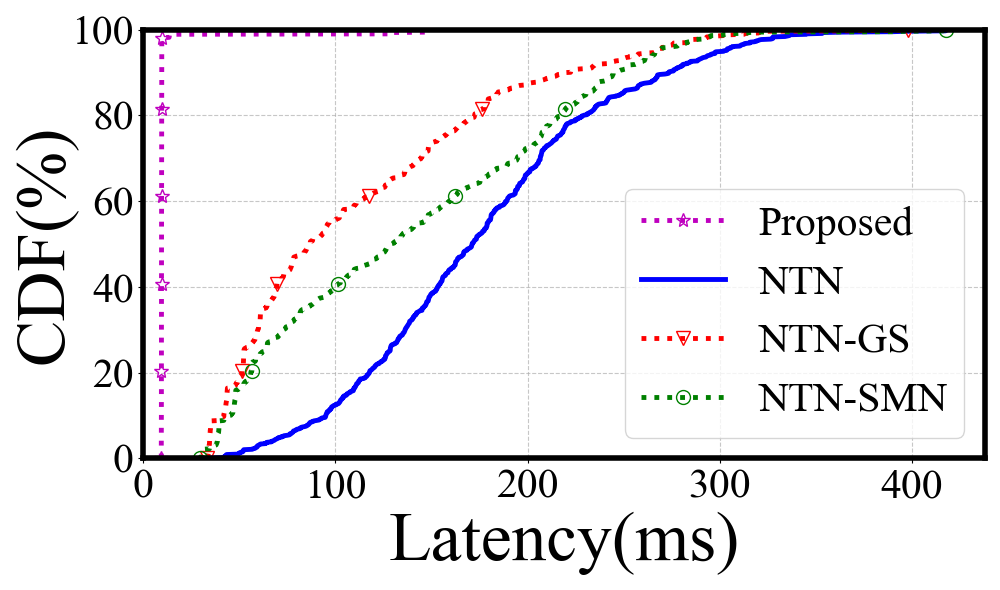}\label{latency_Kuiper}}
\subfloat[]{\includegraphics[width=0.33\textwidth]{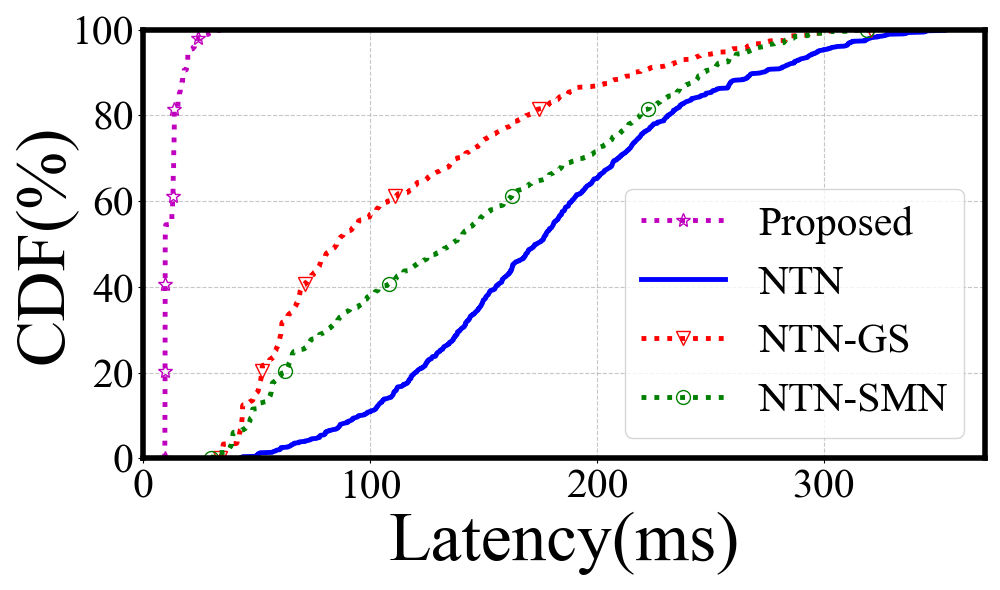}\label{latency_consistent_Kuiper}}
\caption{Comparison of handover latency using different access satellite selection strategies and constellations. 
(a) Flexible strategy, similar direction, Starlink; 
(b) Flexible strategy, all direction, Starlink; 
(c) Consistent strategy, similar direction, Starlink; 
(d) Consistent strategy, all direction, Starlink; 
(e) Flexible strategy, similar direction, Kuiper; 
(f) Consistent strategy, similar direction, Kuiper.}
\label{latency_all}

\end{figure*}

\noindent \textbf{Handover Latency:} As shown in Fig. \ref{latency_all}, we compare the latency of different handover strategies using different access satellite selection strategies and constellations to demonstrate the influence of these metrics. It can be observed that our proposed handover scheme outperforms the other three handover schemes in terms of latency across all scenarios, including both flexible and consistent strategies, as well as two different constellations. More specifically, as shown in Fig.~\ref{latency}, the handover latency based on the proposed handover scheme is 8.8 ms on average, which is much shorter than the handover latency of 250 ms based on the NTN strategy. Meanwhile, it is much shorter than the handover latency based on the two other optimized handover schemes (i.e., NTN-GS and NTN-SMN), which are 153 ms and 158.5 ms, respectively.  

Fig.~\ref{latency} and Fig.~\ref{latency_all_direction} describe the handover latency using different access satellite selection strategies. We can observe that the average handover latency increases by around 6.1$\times$ without optimizing the access satellite selection process compared to the proposed handover scheme, which implies that the simple yet effective optimal access satellite selection scheme proposed in the handover procedure can assist in dramatically reducing the handover latency. This performance gain can be attributed to the transmission delay decrease between satellites when satellites are between the same-direction satellite cluster, since inter-satellite information transfer is unavoidable as S-gNBs need to exchange information. Meanwhile, we also observed similar performance based on other handover strategies, highlighting the necessary requirements for the access satellite selection optimization.

The comparison of Fig.\ref{latency} with Fig.\ref{latency_consistent}, reveals the performance difference of different handover triggering criteria. It can be observed that the handover latency based on the flexible strategy is a little shorter (around 10 ms) than the consistent strategy. The reason is that the source satellite is closer to the target satellite using the flexible strategy as the handover triggering criterion compared to the case of the consistent strategy. However, the flexible strategy brings more handover compared to the consistent strategy. 

Finally, we evaluate the performance of the proposed handover scheme under different satellite constellations such as Starlink and Kuiper. 
First, we can observe the proposed handover strategy outperforms the other three handover schemes irrespective of the kind of satellite constellation, as demonstrated in Fig.~\ref{latency} and Fig.~\ref{latency_Kuiper}. This is because the proposed handover scheme achieves handover latency reduction mainly leveraging the predictable traveling trace of LEO satellites and thus irrelevant to the constellation. Secondly, the handover latency exhibits a little difference (about 10\% on average) under Starlink and Kuiper constellations, which can attributed to the difference in constellation configurations such as inter-orbit and intra-orbit distances and satellite altitudes. Thirdly, we can also observe that there exists a slight handover latency difference when using different access satellite selection strategies in Kuiper satellite constellation, which is verified by comparing Fig.~\ref{latency_Kuiper} with Fig.~\ref{latency_consistent_Kuiper}.

In conclusion, the proposed handover scheme shows a superior performance over the three other handover schemes in terms of handover latency under various conditions.

\begin{figure}[t!]
	\centering
	\includegraphics[width=0.4\textwidth]{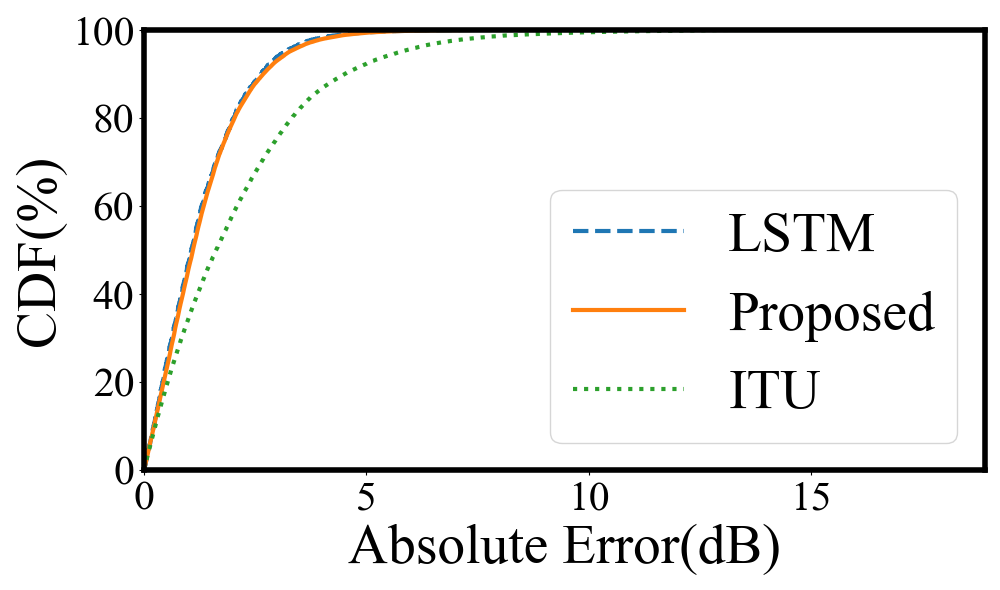}
	\caption{Error of the signal strength prediction}
	\label{signal_strength}
\end{figure}

\begin{table}[t!]
    \centering
    \scriptsize
    \caption{Mean Signal Strength Prediction Error for each satellite-UE pair}
    \begin{tabular}{|l|c|c|c|c|c|c|c|c|}
    \hline
         & \multicolumn{2}{c|}{\textbf{Aqua}} & \multicolumn{2}{c|}{\textbf{Terra}} & \multicolumn{2}{c|}{\textbf{JPSS}} & \multicolumn{2}{c|}{\textbf{NPP}} \\
        \hline
        \textbf{Error(dB)}&\textbf{WI}&\textbf{GM}&\textbf{WI}&\textbf{GM}&\textbf{WI}&\textbf{GM}&\textbf{WI}&\textbf{GM}\\
        \hline
        \textbf{Our} & 1.28 & 0.99 & 1.10 & 1.06 & 1.36 & 1.41 & 1.53 & 1.40 \\
        \hline
        \textbf{LSTM} & 1.19 &1.00& 1.06& 1.08 & 1.33 & 1.39 & 1.42&1.45 \\
        \hline
        \textbf{ITU} & 1.66&1.96& 1.62& 2.09 & 1.88 & 2.48 & 1.95&2.42 \\
        \hline
    \end{tabular}
    
    \label{tab:mean_error}
    \vspace{-0.3cm}
\end{table}

\noindent \textbf{Signal Strength Prediction:} We compare the signal quality prediction accuracy of different algorithms, as shown in Fig. \ref{signal_strength}. Overall, our algorithm demonstrates significantly improved accuracy compared to the channel model proposed by ITU \cite{38811}. The median error for our algorithm is 1.07 dB, and the 90th percentile error is 2.63 dB. In contrast, the ITU model results show median and 90th percentile errors of 1.6 dB and 4.5 dB, respectively. This corresponds to error reductions of 33\% and 43\%, indicating a noticeable improvement over the ITU model. The results indicate that our algorithm outperforms ITU models, particularly in handling long-scale attenuation caused by distance changes. We attribute this improvement to the fact that ITU models are designed for aggregate link behavior and may not capture fine-grained variations induced by the surroundings. For instance, they may not handle local multipath effects or shadow fade caused by the environment around the UE effectively. 

On the other hand, we also compared the performance differences between the proposed method and LSTM. The median error for the LSTM method is 1.06 dB, and the 90th percentile error is 2.59 dB, representing reductions of 1\% and 2\%, respectively, compared to our approach. This shows that with only 1/10 of the parameters as input, the proposed method achieves the accuracy close to that of LSTM. Despite the better performance of LSTM, it is not suitable for our system and is only used as a performance baseline. This is because LSTM predicts future states at fixed intervals based on past sequences, whereas our system requires signal strength predictions at arbitrary time points to ensure accurate handover planning. 

The prediction errors for each satellite-UE pair are shown in Table \ref{tab:mean_error}. For all satellite-UE pairs, our proposed method consistently exhibits smaller average errors, with reductions ranging from 23\% to 50\%. However, errors for certain satellites, such as JPSS and NPP, are significantly larger. One possible reason for this is their higher altitude, which leads to more pronounced effects from the ionosphere.

\begin{figure}[]
	\centering
	\includegraphics[width=0.8\linewidth]{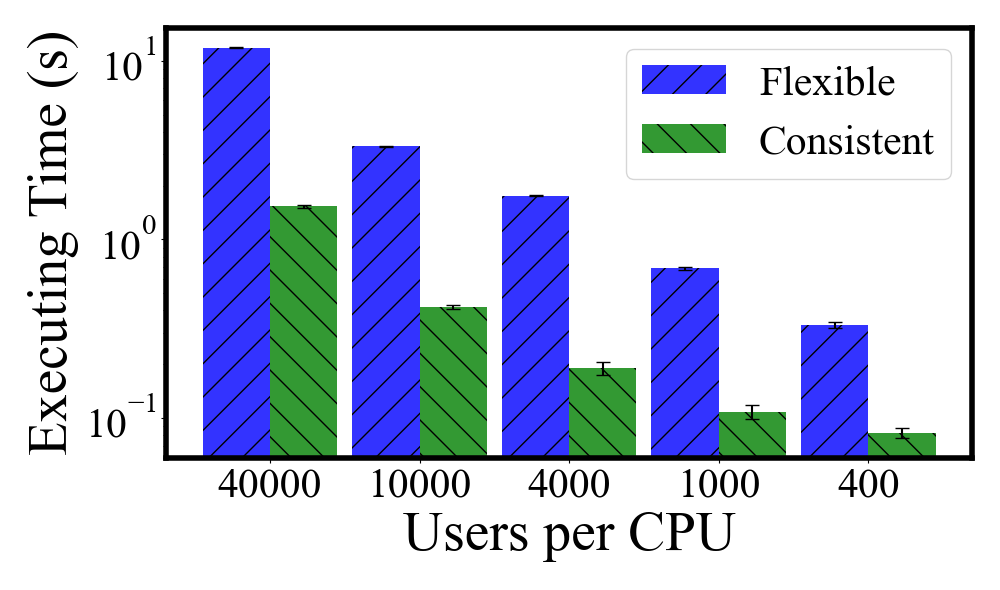}
	\caption{Execution time of handover scheduling algorithm.}
	\label{alogrithm}
\end{figure}

Next, we investigate the execution time of the proposed handover scheduling algorithm with different numbers of users. It can be observed from Fig.~\ref{alogrithm} that the time overhead using the consistent access strategy is much lower than the flexible strategy since there exists much less handover in the case of the consistent access strategy than the flexible strategy. Meanwhile, the results show that the computing time increases with the number of users.

According to the deployment of existing commercial mega-constellations, a satellite serves around 200 users on average \cite{InvestingStarlink}. When the number of users is 200, the computing time is around 0.5 seconds when using a commodity laptop, which is within the time requirements of handover. Even with the predicted increase in the number of users in satellite networks, leading to heightened computing pressure on satellites, the limited satellite-to-ground channel resources may restrict the ability of a single satellite to serve a large user population. Therefore, the actual user load per satellite may not significantly exceed current levels. Considering these factors, we are confident that the computational capability of satellites will remain sufficient to meet the requirements of prediction in the foreseeable future.

\begin{figure}[t!]
    \centering
    \begin{subfigure}{\linewidth}
        \centering
        \includegraphics[width=\linewidth]{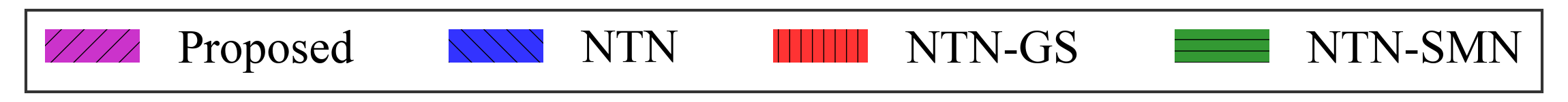}
        \caption*{}  
    \end{subfigure}%
    \vspace{-0.5cm}
    \begin{subfigure}{0.49\linewidth}
        \centering
        \includegraphics[width=\linewidth]{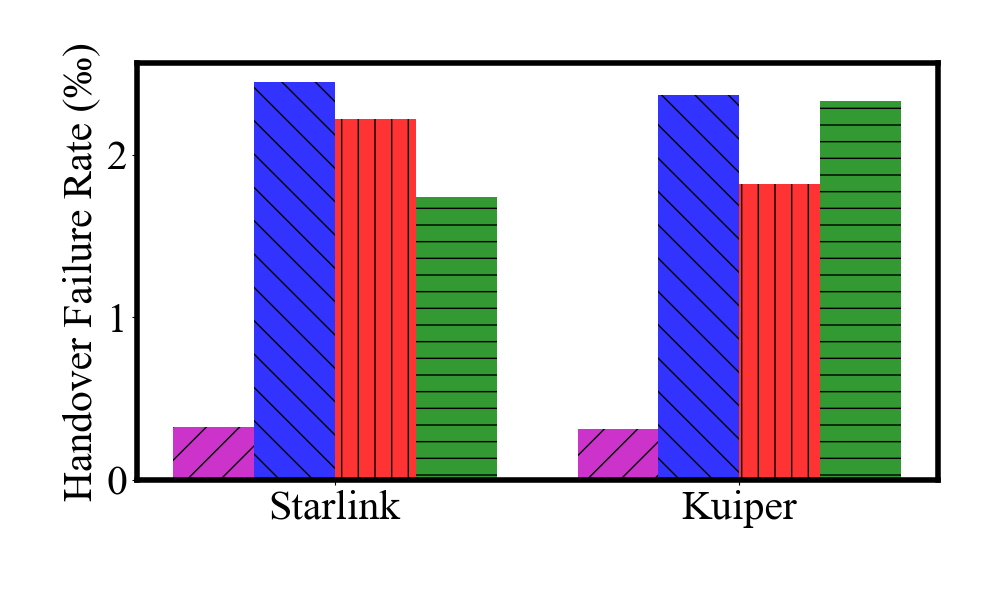}
        \caption{}
        \label{handover failure rate}
    \end{subfigure}
        \hfill
    \begin{subfigure}{0.49\linewidth}
        \centering
        \includegraphics[width=\linewidth]{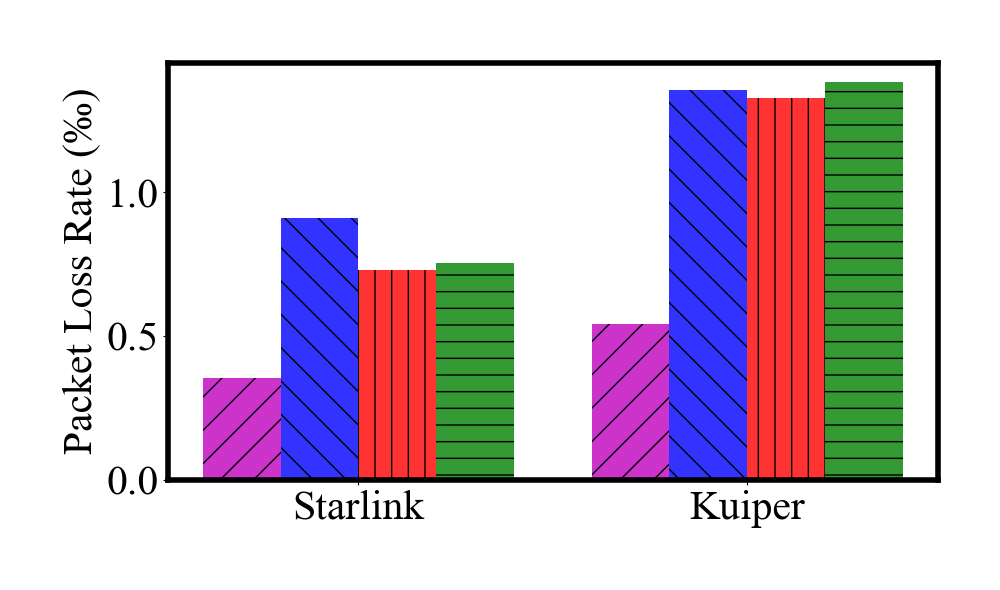}
        \caption{}
        \label{packet loss}
    \end{subfigure}
    \caption{Network Stability. (a) Handover failure rate. (b) Packet loss rate for user-level data.}
    \label{fig:network_performance}
\end{figure}

\noindent \textbf{Network Stability:} Next, we discuss the impact of handover duration in the proposed scheme on network stability.

On one hand, the handover failure rate is influenced by the handover duration. If the user moves out of the target satellite’s service area due to satellite mobility during the handover process, the handover will fail. We evaluated the handover failure rates under different schemes, as shown in Fig. \ref{handover failure rate}. Specifically, the proposed scheme achieves a handover failure rate of approximately 0.3\textperthousand~whereas the compared methods exhibit failure rates between 1.5\textperthousand~and 2.5\textperthousand~in Starlink constellation, and showing a similar pattern in Kuiper. The primary reason for this improvement is that the proposed approach significantly reduces handover latency, thereby decreasing the likelihood of failure. It is also worth noting that, in the absence of packet loss, the overall handover failure probability remains low. This is because this study does not consider packet loss caused by channel quality, interference, or satellite mobility, as these factors are independent of the handover process itself and beyond the scope of this discussion. 

On the other hand, the packet loss rate of user-level data is also related to handover duration. If a user initiates a handover while data transmission is in progress, some transmitted data may be lost. This implies that a shorter handover duration contributes to higher network stability. We compared the impact of different schemes on packet loss rates, as shown in Fig. \ref{packet loss}. Specifically, the proposed scheme achieves a packet loss rate of approximately 0.35\textperthousand, while the compared methods exhibit packet loss rates between 0.7 \textperthousand~and 0.9\textperthousand~in Starlink constellation. The proposed scheme reduces the handover failure rate by more than 2 times.

\begin{figure}[t!]
    \vspace{-0.3cm}
\centering
   \subfloat[]{\includegraphics[height=0.16\textwidth]{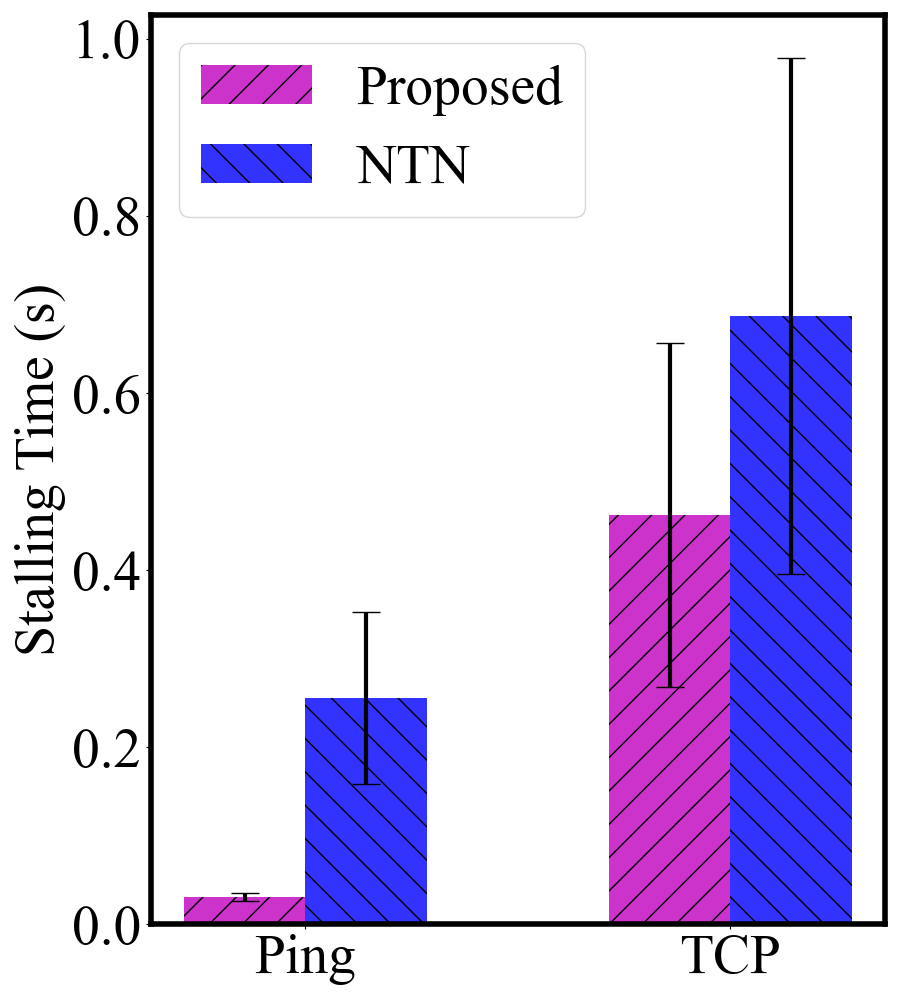}\label{sta}}
    \subfloat[]{\includegraphics[height=0.16\textwidth]{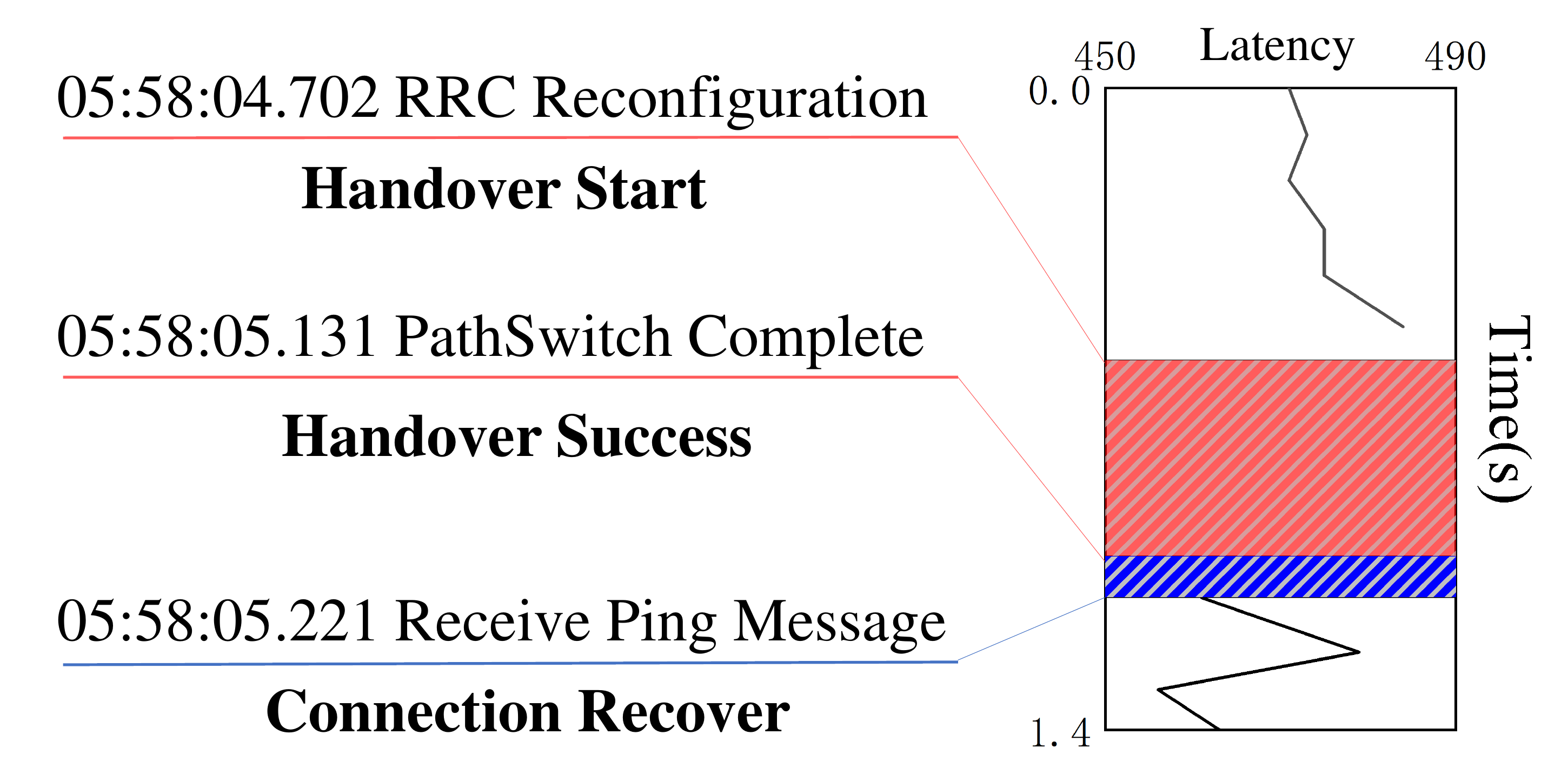}\label{ping detail}}
    \caption{User level performance. (a) Stalling Time. (b) Ping in NTN.}\label{User level}
    \vspace{-0.6cm}
\end{figure}
\noindent \textbf{User-level performance:} Furthermore, we have compared the user-level performance respectively brought by the proposed handover scheme and 5G NTN handover strategy. As shown in Fig. \ref{User level}, the stalling time based on the proposed handover scheme can be reduced by 89\% compared to 5G NTN handover strategy. Meanwhile, in the case of TCP flows, our proposed handover scheme can decrease the stalling time by 33\% compared to 5G NTN handover, as demonstrated in Fig. \ref{sta}. This is because handover affects the user-level performance, where large handover latency prolongs the delay of user-level applications and then deteriorates the overall performance.   

In the meantime, the reason behind the limited performance improvement in the case of TCP flows is that there exists a three-way handshake mechanism incurring substantial time overhead, which is exacerbated by the long propagation latency between inter-satellites. To illustrate in detail, we delve into details of the ping procedure in mobile satellite networks. As shown in Fig.~\ref{ping detail}, the stalling time in the link recovery process can be divided into two steps. First, the user terminal connects with the target S-gNB. Secondly, after a certain duration, the connection between the user and the server is restored. The proposed handover scheme can considerably reduce the time delay of the first step, i.e., the time taken for the link to recover. However, the overhead of the second step is primarily caused by the delay between inter-satellites, which is beyond the scope of this paper and thus results in limited performance improvement in terms of stalling time.

\begin{figure}[]
	\centering
	\includegraphics[width=\linewidth]{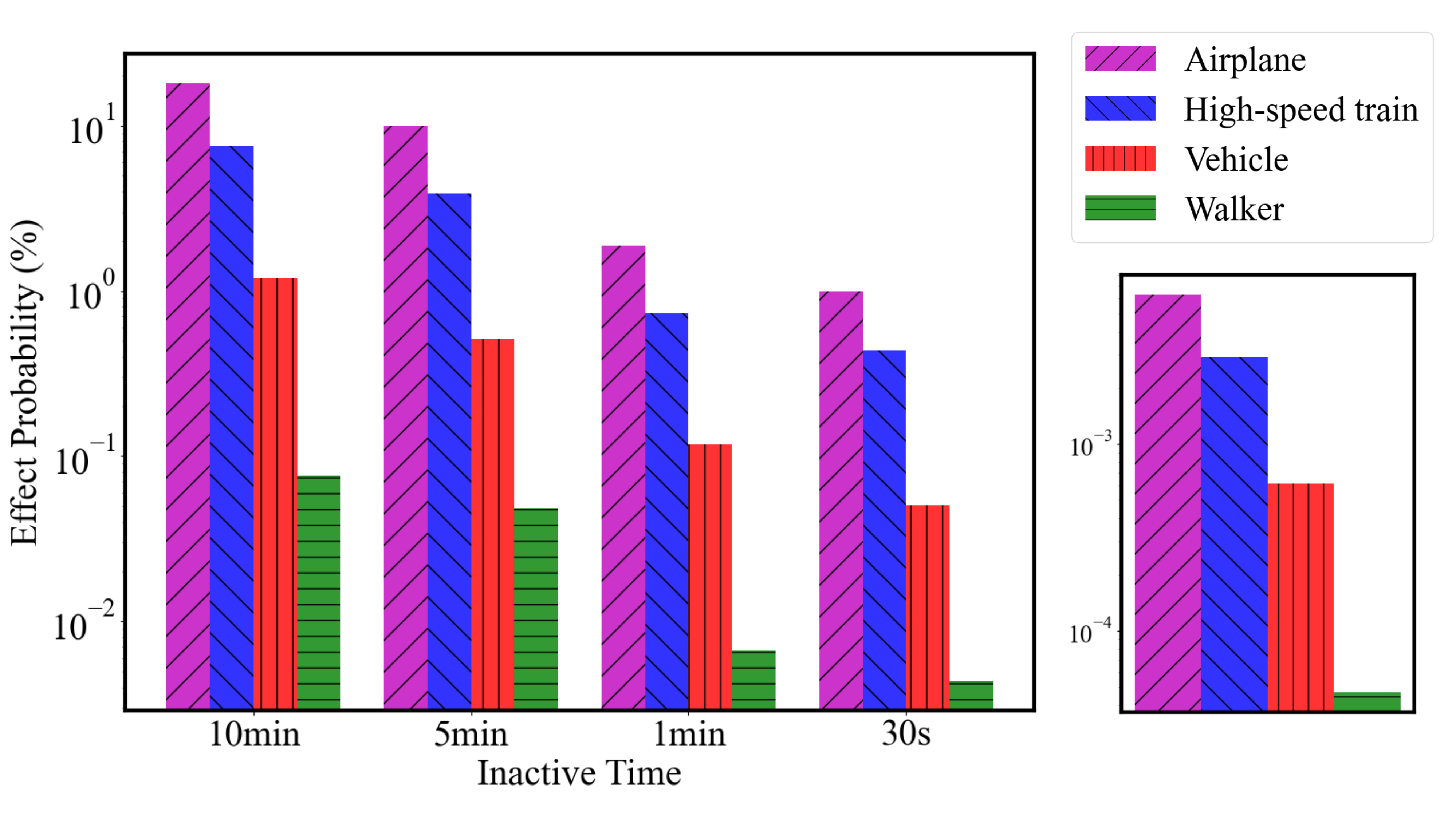}
	\caption{Impact of user mobility on inadequate scheduling. (left: user is inactive; right: user is active.)}\label{fail}
\end{figure}

\noindent \textbf{Impact of User Movement:} Finally, we conduct experiments to evaluate the influence of user movement (e.g., users move at different speeds in both active and inactive scenarios) on handover performance, referred to as ‘inadequate handover scheduling’ defined in §\ref{users_movement_fail}. As shown in Fig. \ref{fail}, for active ground users, the probability of 'inappropriate handover scheduling' is very low, approximately $10^{-5}$ for users in high-speed airplanes and lower in other scenarios. In the case of walking speed, the abnormal probability is in the order of a millionth, which can be considered negligible.

On the other hand, for users who are in high-speed motion and remain inactive for an extended period, the probability of triggering an 'inadequate handover scheduling' is relatively higher. For example, after 10 minutes of inactivity, users on airplanes have an 18\% probability of experiencing an abnormal handover, while users travelling on the high-speed train have a 7\% abnormal probability.

\section{Related Work}
The problem of handover in satellite networks has been extensively discussed. Considering the satellite network's topology is a natural approach \cite{8926396,9003306}. One mainstream method is the virtual node method \cite{8926396}, in which partition mappings satellites and ground stations into geographic regions. However, this method only works in polar orbit constellation networks and is not applicable to mainstream constellation configurations, such as the Walker constellation primarily used by Starlink. Another mainstream method is based on calculating all possible network topology states to predict the network topology at any given moment \cite{9003306}. However, this method's computational complexity and storage requirements increase exponentially with the number of satellites and users, making it impractical for large-scale satellite networks.

Another approach is modifying the network structure on the basis of standard 5G network \cite{stateless_mobile,9755268,9367419,10121544, handover_strategy,guptaai,channelawarehandover,SEHAP}. This includes introducing new NFs and leveraging advanced 5G handover procedures. However, most of these methods do not specifically prioritize handover latency optimization or provide sufficient improvements in reducing latency. A recent work \cite{stateless_mobile} introduced stateless network elements on the satellite side, allowing users to store the information required by the core network, thus avoiding interactions with the core network during handover switches. 
However, this work primarily focused on the core network and neglected RAN, which could lead to an underestimation of the overall latency in the handover procedure.
Regarding the handover procedure, \cite{guptaai} focuses on directly applying machine learning for handover prediction, while \cite{channelawarehandover} proposes making handover decisions based on prediction results rather than predefined policies. However, both studies concentrate solely on modifications to the RAN side of the handover process, without considering interactions with the core network, which is the dominant contributor to latency in satellite networks.


Our proposed scheme involves predicting the network topology (i.e., the user's access satellite). However, there is no need to compute the global topology; instead, each UPF calculates the users it is responsible for respectively, significantly reducing the computational load. 
On the other hand, we modify the handover procedure. By completely eliminating the interaction with core network during handover, we achieve a substantial reduction in latency, surpassing existing methods.  Our solution takes both core network and RAN into consideration, and we conducted comprehensive system-level experiments on a latency platform. This ensures that experimental results are close to real-world scenarios.

\section{Discussion and limitation}
{\textbf{Practicability}: The proposed method demonstrates strong practicability and is deployable in future satellite mobile networks. Specifically, it only requires the introduction of the SSF into the core network and a slight modification to the gNB onboard satellites to support the proposed path handover mechanism. No changes to the UE or existing network functions are needed. We have implemented a prototype based on this architecture, which validates the practical feasibility of the proposed solution.}

{\textbf{Extending to Larger-Scale Satellite Constellations}: We conduct our experiments using commercial constellations such as Starlink and Kuiper, both of which are Walker-structured systems comprising thousands of satellites. Notably, the proposed design is not constrained by any specific constellation architecture or scale. Its core advantage lies in leveraging the predictability of satellite orbits—once the orbital parameters are known (typically obtainable from regulatory agencies or constellation operators), the algorithm can operate effectively regardless of constellation size. Although the computational complexity of the prediction process naturally grows with the number of satellites, this challenge is mitigated by the algorithm's inherent support for parallelization. By scaling the computational infrastructure, real-time performance can be maintained with a relatively modest increase in cost compared to the overall investment in the satellite network. }

\textbf{Access Satellite Selection}:
In this paper, we have evaluated the performance of different kinds of access satellite selection strategies. We find that the consistent strategy can considerably reduce the number of handover while the flexible strategy can provide higher-quality service yet result in more handover. It still remains an open problem whether these two access satellite selection strategies or another option is suitable for the mobile satellite network, which is beyond the scope of the research work in this paper. We believe the access satellite selection scheme is highly related to the applications and leave it for future work.

\textbf{Access Satellite of Ground Station}: In this paper, we have only discussed the update process of access satellites. However, changes can also occur for the access satellites corresponding to ground stations. Due to the relatively small number and fixed positions of ground stations, satellites can adapt to these changes through updates in their own routing strategies, without the need for reflection in the mobile network's signaling procedures.
%

\section{Conclusion}
Frequent handover resulting from the fast traveling speed of LEO satellites is a significant challenge in mobile satellite networks. In this paper, we propose a novel handover design to reduce handover latency. Based on predicting user access satellites using ephemeris, we avoid communication with core network. We constructed our mobile satellite network prototype using Open5GS and UERANSIM and conducted experiments to assess the performance of the proposed design. The results show that our approach achieved a remarkable 10$\times$ reduction in latency compared to the standard NTN handover scheme and two other existing approaches.  As a potential future direction, we are looking forward to extending our Hurry to improve the performance of various applications such as distributed learning systems~\cite{lin2024adaptsfl,lin2025hierarchical,hu2024accelerating,zhang2024fedac,lin2024efficient,chen2021rf,lyu2023optimal,lin2025hasfl,zhang2025lcfed,lin2024split}, large language models~\cite{kasneci2023chatgpt,fang2024automated,lin2025hsplitlora,qiu2024ifvit,jin2024large,lin2023pushing,yang2024give,lin2024splitlora,chang2024survey}, etc in LEO satellite networks.

\section{Acknowledgments}
This work has been supported by the National Natural Science Foundation of China Grant No. 62341105.

\bibliographystyle{IEEEtran}
\bibliography{ref.bib}

\begin{thebibliography}{10}
\providecommand{\url}[1]{#1}
\csname url@samestyle\endcsname
\providecommand{\newblock}{\relax}
\providecommand{\bibinfo}[2]{#2}
\providecommand{\BIBentrySTDinterwordspacing}{\spaceskip=0pt\relax}
\providecommand{\BIBentryALTinterwordstretchfactor}{4}
\providecommand{\BIBentryALTinterwordspacing}{\spaceskip=\fontdimen2\font plus
\BIBentryALTinterwordstretchfactor\fontdimen3\font minus \fontdimen4\font\relax}
\providecommand{\BIBforeignlanguage}[2]{{%
\expandafter\ifx\csname l@#1\endcsname\relax
\typeout{** WARNING: IEEEtran.bst: No hyphenation pattern has been}%
\typeout{** loaded for the language `#1'. Using the pattern for}%
\typeout{** the default language instead.}%
\else
\language=\csname l@#1\endcsname
\fi
#2}}
\providecommand{\BIBdecl}{\relax}
\BIBdecl

\bibitem{infocom24handover}
J.~Wu, S.~Su, X.~Wang, J.~Zhang, and Y.~Gao, ``Accelerating handover in mobile satellite network,'' in \emph{IEEE INFOCOM 2024 - IEEE Conference on Computer Communications}, 2024, pp. 531--540.

\bibitem{starlink}
``Starlink,'' \url{https://www.starlink.com/}.

\bibitem{lin2024fedsn}
Z.~Lin, Z.~Chen, Z.~Fang, X.~Chen, X.~Wang, and Y.~Gao, ``Fedsn: A federated learning framework over heterogeneous leo satellite networks,'' \emph{IEEE Transactions on Mobile Computing}, 2024.

\bibitem{peng2025sigchord}
J.~Peng, J.~Duan, Z.~Lin, H.~Yuan, Y.~Gao, and Z.~Chen, ``{SigChord: Sniffing Wide Non-Sparse Multiband Signals for Terrestrial and Non-Terrestrial Wireless Networks},'' \emph{arXiv preprint arXiv:2504.06587}, 2025.

\bibitem{yuan2024satsense}
H.~Yuan, Z.~Chen, Z.~Lin, J.~Peng, Z.~Fang, Y.~Zhong, Z.~Song, and Y.~Gao, ``{SatSense: Multi-Satellite Collaborative Framework for Spectrum Sensing},'' \emph{{IEEE} Trans. Cogn. Commun. Netw.}, 2025.

\bibitem{zhao2024leo}
Z.~Zhao, Z.~Chen, Z.~Lin, W.~Zhu, K.~Qiu, C.~You, and Y.~Gao, ``{LEO Satellite Networks Assisted Geo-Distributed Data Processing},'' \emph{{IEEE} Wireless Commun. Lett.}, 2024.

\bibitem{lin2025leo}
Z.~Lin, Y.~Zhang, Z.~Chen, Z.~Fang, C.~Wu, X.~Chen, Y.~Gao, and J.~Luo, ``{LEO-Split: A Semi-Supervised Split Learning Framework over LEO Satellite Networks},'' \emph{arXiv preprint arXiv:2501.01293}, 2025.

\bibitem{yuan2023graph}
H.~Yuan, Z.~Chen, Z.~Lin, J.~Peng, Z.~Fang, Y.~Zhong, Z.~Song, X.~Wang, and Y.~Gao, ``{Graph Learning for Multi-Satellite Based Spectrum Sensing},'' in \emph{{Proc. IEEE Int. Conf. Commun. Technol. (ICCT)}}, 2023, pp. 1112--1116.

\bibitem{peng2024sums}
J.~Peng, Z.~Chen, Z.~Lin, H.~Yuan, Z.~Fang, L.~Bao, Z.~Song, Y.~Li, J.~Ren, and Y.~Gao, ``{SUMS: Sniffing Unknown Multiband Signals under Low Sampling Rates},'' \emph{{IEEE} Trans. Mobile Comput.}, 2024.

\bibitem{zhang2024satfed}
Y.~Zhang, Z.~Lin, Z.~Chen, Z.~Fang, W.~Zhu, X.~Chen, J.~Zhao, and Y.~Gao, ``Satfed: A resource-efficient leo satellite-assisted heterogeneous federated learning framework,'' \emph{arXiv preprint arXiv:2409.13503}, 2024.

\bibitem{yuan2025constructing}
H.~Yuan, Z.~Chen, Z.~Lin, J.~Peng, Y.~Zhong, X.~Hu, S.~Xue, W.~Li, and Y.~Gao, ``{Constructing 4D Radio Map in LEO Satellite Networks with Limited Samples},'' \emph{{IEEE} INFOCOM}, 2025.

\bibitem{23501}
3GPP, ``{TS 23.501,System Architecture for the {5G} System},'' 2023.

\bibitem{23502}
------, ``{TS 23.502,procedures for the 5G system},'' 2023.

\bibitem{6Gwhite}
``{White Paper on 6G Vision and Candidate Technologies},'' IMT-2030 (6G) Promotion Group, Tech. Rep., 2018.

\bibitem{t-mobile}
\BIBentryALTinterwordspacing
``{T‑Mobile takes coverage above and beyond with SpaceX}.'' [Online]. Available: \url{https://www.t-mobile.com/news/un-carrier/t-mobile-takes-coverage-above-and-beyond-with-spacex}
\BIBentrySTDinterwordspacing

\bibitem{optus}
\BIBentryALTinterwordspacing
``{Optus and Elon Musk’s Starlink to Offer Mobile Connectivity via Satellite}.'' [Online]. Available: \url{https://gizmodo.com.au/2023/07/optus-starlink-mobile-connectivity-satellite/}
\BIBentrySTDinterwordspacing

\bibitem{38821}
3GPP, ``{TS 38.821,solutions for NR to support non-terrestrial networks (NTN)},'' 2024.

\bibitem{stateless_mobile}
Y.~Li, H.~Li, W.~Liu, L.~Liu, Y.~Chen, J.~Wu, Q.~Wu, J.~Liu, and Z.~Lai, ``A case for stateless mobile core network functions in space,'' in \emph{Proceeding of ACM SIGCOMM}.\hskip 1em plus 0.5em minus 0.4em\relax Association for Computing Machinery, 2022, p. 298–313.

\bibitem{re-test}
\BIBentryALTinterwordspacing
TTP, ``{TTP develops 5G NTN test environment with Keysight tools}.'' [Online]. Available: \url{https://www.ttp.com/news/ttp-develops-5g-ntn-test-environment-with-keysight-tools/}
\BIBentrySTDinterwordspacing

\bibitem{Kuiper}
``Project kuiper,'' \url{https://www.aboutamazon.com/what-we-do/devices-services/project-kuiper}.

\bibitem{InvestingStarlink}
{CNBC}, ``{Investing in Space: Is SpaceX’s Starlink growing satellite internet market share, or taking it?}'' \url{https://cnb.cx/42Pa8jd}, Mar 2023.

\bibitem{9393372}
A.~U. Chaudhry and H.~Yanikomeroglu, ``Laser intersatellite links in a {Starlink} constellation: A classification and analysis,'' \emph{IEEE Vehicular Technology Magazine}, vol.~16, no.~2, pp. 48--56, 2021.

\bibitem{Sateliot}
\BIBentryALTinterwordspacing
``{Sateliot | Space · Connecting · 5G Satellite IoT}.'' [Online]. Available: \url{https://sateliot.space/en/}
\BIBentrySTDinterwordspacing

\bibitem{38811}
3GPP, ``{TS 38811, study on new radio (NR) to support non-terrestrial networks},'' 2023.

\bibitem{4062836}
P.~K. Chowdhury, M.~Atiquzzaman, and W.~Ivancic, ``{Handover chemes in satellite networks: state-of-the-art and future research directions},'' \emph{IEEE Communications Surveys \& Tutorials}, vol.~8, no.~4, pp. 2--14, 2006.

\bibitem{Satellite_handover}
E.~Papapetrou, S.~Karapantazis, G.~Dimitriadis, and F.-N. Pavlidou, ``{Satellite handover techniques for LEO networks},'' \emph{International Journal of Satellite Communications and Networking}, vol.~22, no.~2, pp. 231--245, 2004.

\bibitem{neinavaie2022unveiling}
M.~Neinavaie and Z.~M. Kassas, ``Unveiling beamforming strategies of starlink {LEO} satellites,'' in \emph{Proceedings of the 35th International Technical Meeting of the Satellite Division of The Institute of Navigation}, 2022, pp. 2525--2531.

\bibitem{oneweb}
J.~Radtke, C.~Kebschull, and E.~Stoll, ``Interactions of the space debris environment with mega constellations—using the example of the oneweb constellation,'' \emph{Acta Astronautica}, vol. 131, pp. 55--68, 2017.

\bibitem{chen2021analysis}
Q.~Chen, G.~Giambene, L.~Yang, C.~Fan, and X.~Chen, ``{Analysis of inter-satellite link paths for LEO mega-constellation networks},'' \emph{IEEE Transactions on Vehicular Technology}, vol.~70, no.~3, pp. 2743--2755, 2021.

\bibitem{zhang2022enabling}
Y.~Zhang, Q.~Wu, Z.~Lai, and H.~Li, ``{Enabling low-latency-capable satellite-ground topology for emerging LEO satellite networks},'' in \emph{Proceeding of IEEE INFOCOM}.\hskip 1em plus 0.5em minus 0.4em\relax IEEE, 2022, pp. 1329--1338.

\bibitem{orbitprediction}
J.~Haidar-Ahmad, N.~Khairallah, and Z.~M. Kassas, ``A hybrid analytical-machine learning approach for {LEO} satellite orbit prediction,'' in \emph{Proceeding of 25th International Conference on Information Fusion}, 2022, pp. 1--7.

\bibitem{rs15010133}
\BIBentryALTinterwordspacing
K.~Wang, J.~Liu, H.~Su, A.~El-Mowafy, and X.~Yang, ``Real-time {LEO} satellite orbits based on batch least-squares orbit determination with short-term orbit prediction,'' \emph{Remote Sensing}, vol.~15, no.~1, 2023. [Online]. Available: \url{https://www.mdpi.com/2072-4292/15/1/133}
\BIBentrySTDinterwordspacing

\bibitem{celestrak}
``{Celestrak},'' \url{https://celestrak.com/}.

\bibitem{StarlinkStatus}
``{Starlink Status},'' \url{https://starlinkstatus.space/}.

\bibitem{skyfield}
``Skyfield - documentation,'' \url{https://rhodesmill.org/skyfield/}.

\bibitem{L2D2}
D.~Vasisht, J.~Shenoy, and R.~Chandra, ``{L2D2}: Low latency distributed downlink for {LEO} satellites,'' in \emph{Proceedings of the 2021 ACM SIGCOMM 2021 Conference}, 2021, pp. 151--164.

\bibitem{ueransim}
``{UERANSIM},'' \url{https://github.com/aligungr/UERANSIM/}.

\bibitem{open5gs}
``{open5gs.org},'' \url{https://open5gs.org/}.

\bibitem{dong2018mianxiang}
D.~Yanlei, L.~Dong'ang, L.~Qin, W.~Chunting, and S.~Keyi, ``{Research on mobility management strategy based on the mobile foreign agent domain in satellite networks},'' \emph{Journal of Xidian University}, vol.~45, no. 156-162, 2018.

\bibitem{7811041}
W.~Han, B.~Wang, Z.~Feng, B.~Zhao, and W.~Yu, ``{Distributed mobility management in IP/LEO satellite networks},'' in \emph{International Conference on Systems and Informatics}, 2016, pp. 691--695.

\bibitem{9755268}
J.~Kim, J.~Lee, H.~Ko, T.~Kim, and S.~Pack, ``Space mobile networks: Satellite as core and access networks for {B5G},'' \emph{IEEE Communications Magazine}, vol.~60, no.~4, pp. 58--64, 2022.

\bibitem{lstm}
\BIBentryALTinterwordspacing
S.~Hochreiter and J.~Schmidhuber, ``Long short-term memory,'' \emph{Neural Comput.}, vol.~9, no.~8, p. 1735–1780, Nov. 1997. [Online]. Available: \url{https://doi.org/10.1162/neco.1997.9.8.1735}
\BIBentrySTDinterwordspacing

\bibitem{8926396}
Q.~Chen, J.~Guo, L.~Yang, X.~Liu, and X.~Chen, ``Topology virtualization and dynamics shielding method for {LEO} satellite networks,'' \emph{IEEE Communications Letters}, vol.~24, no.~2, pp. 433--437, 2020.

\bibitem{9003306}
T.~Zhang, J.~Li, H.~Li, S.~Zhang, P.~Wang, and H.~Shen, ``Application of time-varying graph theory over the space information networks,'' \emph{IEEE Network}, vol.~34, no.~2, pp. 179--185, 2020.

\bibitem{9367419}
E.~Juan, M.~Lauridsen, J.~Wigard, and P.~E. Mogensen, ``{5G New Radio mobility performance in LEO-based non-Terrestrial networks},'' in \emph{Proceeding of IEEE Globecom Workshops}, 2020, pp. 1--6.

\bibitem{10121544}
J.~Wu, Y.~Gao, L.~Wang, J.~Zhang, and D.~O. Wu, ``How to allocate resources in cloud native networks towards {6G},'' \emph{IEEE Network}, pp. 1--7, 2023.

\bibitem{handover_strategy}
S.~Eydian, M.~Hosseini, and G.~Karabulut~Kurt, ``Handover strategy for {LEO} satellite networks using bipartite graph and hysteresis margin,'' \emph{IEEE Open Journal of the Communications Society}, vol.~6, pp. 1470--1484, 2025.

\bibitem{guptaai}
H.~Gupta, N.~Srivastava, and L.~Borman, ``{AI}-based handover decision algorithm for conditional handover in non-terrestrial networks ({NTNs}),'' \emph{2025 Workshop on Computing, Networking and Communications {(CNC)}}, 2025.

\bibitem{channelawarehandover}
C.~Nuo, S.~Zhili, S.~Yujie, C.~Yue, X.~Xu, and A.~B. Sali, ``Channel-aware handover management for space-air-ground integrated networks,'' \emph{China Communications}, vol.~22, no.~2, pp. 62--76, 2025.

\bibitem{SEHAP}
Y.~Guo, J.~Wang, K.~Geng, Z.~Li, F.~Li, and L.~Fang, ``{SEHAP}: Secure and efficient handover authentication protocol in {LEO} satellite non-terrestrial networks,'' in \emph{ICASSP 2025 - 2025 IEEE International Conference on Acoustics, Speech and Signal Processing (ICASSP)}, 2025, pp. 1--5.

\bibitem{lin2024adaptsfl}
Z.~Lin, G.~Qu, W.~Wei, X.~Chen, and K.~K. Leung, ``{Adaptsfl: Adaptive Split Federated Learning in Resource-Constrained Edge Networks},'' \emph{{IEEE} Trans. Netw.}, 2024.

\bibitem{lin2025hierarchical}
Z.~Lin, W.~Wei, Z.~Chen, C.-T. Lam, X.~Chen, Y.~Gao, and J.~Luo, ``{Hierarchical Split Federated Learning: Convergence Analysis and System Optimization},'' \emph{{IEEE} Trans. Mobile Comput.}, 2025.

\bibitem{hu2024accelerating}
M.~Hu, J.~Zhang, X.~Wang, S.~Liu, and Z.~Lin, ``{Accelerating Federated Learning with Model Segmentation for Edge Networks},'' \emph{{IEEE} Trans. Green Commun. Netw.}, 2024.

\bibitem{zhang2024fedac}
Y.~Zhang, H.~Chen, Z.~Lin, Z.~Chen, and J.~Zhao, ``Fedac: An adaptive clustered federated learning framework for heterogeneous data,'' \emph{arXiv preprint arXiv:2403.16460}, 2024.

\bibitem{lin2024efficient}
Z.~Lin, G.~Zhu, Y.~Deng, X.~Chen, Y.~Gao, K.~Huang, and Y.~Fang, ``{Efficient Parallel Split Learning over Resource-Constrained Wireless Edge Networks},'' \emph{{IEEE} Trans. Mobile Comput.}, vol.~23, no.~10, pp. 9224--9239, 2024.

\bibitem{chen2021rf}
Z.~Chen, C.~Cai, T.~Zheng, J.~Luo, J.~Xiong, and X.~Wang, ``Rf-based human activity recognition using signal adapted convolutional neural network,'' \emph{IEEE Transactions on Mobile Computing}, vol.~22, no.~1, pp. 487--499, 2021.

\bibitem{lyu2023optimal}
S.~Lyu, Z.~Lin, G.~Qu, X.~Chen, X.~Huang, and P.~Li, ``Optimal resource allocation for u-shaped parallel split learning,'' in \emph{2023 IEEE Globecom Workshops (GC Wkshps)}, 2023, pp. 197--202.

\bibitem{lin2025hasfl}
Z.~Lin, Z.~Chen, X.~Chen, W.~Ni, and Y.~Gao, ``{HASFL: Heterogeneity-Aware Split Federated Learning over Edge Computing Systems},'' \emph{arXiv preprint arXiv:2506.08426}, 2025.

\bibitem{zhang2025lcfed}
Y.~Zhang, H.~Chen, Z.~Lin, Z.~Chen, and J.~Zhao, ``{LCFed: An Efficient Clustered Federated Learning Framework for Heterogeneous Data},'' \emph{arXiv preprint arXiv:2501.01850}, 2025.

\bibitem{lin2024split}
Z.~Lin, G.~Qu, X.~Chen, and K.~Huang, ``{Split Learning in 6G Edge Networks},'' \emph{{IEEE} Wirel. Commun.}, 2024.

\bibitem{kasneci2023chatgpt}
E.~Kasneci, K.~Se{\ss}ler, S.~K{\"u}chemann, M.~Bannert, D.~Dementieva, F.~Fischer, U.~Gasser, G.~Groh, S.~G{\"u}nnemann, E.~H{\"u}llermeier \emph{et~al.}, ``Chatgpt for good? on opportunities and challenges of large language models for education,'' \emph{Learning and individual differences}, vol. 103, p. 102274, 2023.

\bibitem{fang2024automated}
Z.~Fang, Z.~Lin, Z.~Chen, X.~Chen, Y.~Gao, and Y.~Fang, ``{Automated Federated Pipeline for Parameter-Efficient Fine-Tuning of Large Language Models},'' \emph{arXiv preprint arXiv:2404.06448}, 2024.

\bibitem{lin2025hsplitlora}
Z.~Lin, Y.~Zhang, Z.~Chen, Z.~Fang, X.~Chen, P.~Vepakomma, W.~Ni, J.~Luo, and Y.~Gao, ``{HSplitLoRA: A Heterogeneous Split Parameter-Efficient Fine-Tuning Framework for Large Language Models},'' \emph{arXiv preprint arXiv:2505.02795}, 2025.

\bibitem{qiu2024ifvit}
Y.~Qiu, H.~Chen, X.~Dong, Z.~Lin, I.~Y. Liao, M.~Tistarelli, and Z.~Jin, ``Ifvit: Interpretable fixed-length representation for fingerprint matching via vision transformer,'' \emph{IEEE Transactions on Information Forensics and Security}, 2024.

\bibitem{jin2024large}
B.~Jin, G.~Liu, C.~Han, M.~Jiang, H.~Ji, and J.~Han, ``Large language models on graphs: A comprehensive survey,'' \emph{IEEE Transactions on Knowledge and Data Engineering}, 2024.

\bibitem{lin2023pushing}
Z.~Lin, G.~Qu, Q.~Chen, X.~Chen, Z.~Chen, and K.~Huang, ``{Pushing Large Language Models to the 6G Edge: Vision, Challenges, and Opportunities},'' \emph{arXiv preprint arXiv:2309.16739}, 2023.

\bibitem{yang2024give}
L.~Yang, H.~Chen, Z.~Li, X.~Ding, and X.~Wu, ``Give us the facts: Enhancing large language models with knowledge graphs for fact-aware language modeling,'' \emph{IEEE Transactions on Knowledge and Data Engineering}, vol.~36, no.~7, pp. 3091--3110, 2024.

\bibitem{lin2024splitlora}
Z.~Lin, X.~Hu, Y.~Zhang, Z.~Chen, Z.~Fang, X.~Chen, A.~Li, P.~Vepakomma, and Y.~Gao, ``{SplitLoRA: A Split Parameter-Efficient Fine-Tuning Framework for Large Language Models},'' \emph{arXiv preprint arXiv:2407.00952}, 2024.

\bibitem{chang2024survey}
Y.~Chang, X.~Wang, J.~Wang, Y.~Wu, L.~Yang, K.~Zhu, H.~Chen, X.~Yi, C.~Wang, Y.~Wang \emph{et~al.}, ``A survey on evaluation of large language models,'' \emph{ACM transactions on intelligent systems and technology}, vol.~15, no.~3, pp. 1--45, 2024.

\end{thebibliography}

\end{document}